\documentstyle[aps,prd,eqsecnum,graphicx]{revtex}

\begin{document}

\draft

\title{Quark mixings and flavor changing interactions
\protect \\
with singlet quarks}

\author{Katsuichi Higuchi and Katsuji Yamamoto}

\address{Department of Nuclear Engineering,
Kyoto University, Kyoto 606-8501, Japan}


\maketitle

\begin{abstract}
Aspects of the quark mixings and flavor changing interactions
are investigated in electroweak models with singlet quarks.
The effects on the ordinary quark mixing are determined
in terms of the quark masses and the parameters describing
the mixing between the ordinary quarks $ q $ and the singlet quarks $ Q $
($ q $-$ Q $ mixing).
Some salient features arise in the flavor changing interactions
through the $ q $-$ Q $ mixing.
The unitarity of the Cabibbo-Kobayashi-Maskawa (CKM) matrix
within the ordinary quark sector is violated,
and the flavor changing neutral currents (FCNC's) appear
both in the gauge and scalar couplings.
The flavor changing interactions are calculated appropriately
in terms of the $ q $-$ Q $ mixing parameters and the quark masses,
which really exhibit specific flavor structures.
It is found that there are reasonable ranges of the model parameters
to reproduce the ordinary quark mass hierarchy and the actual CKM structure
even in the presence of $ q $-$ Q $ mixing.
Some phenomenological effects of the singlet quarks are also discussed.
In particular, the scalar FCNC's
may be more important in some cases,
if the singlet quarks as well as the extra scalar particles
from the singlet Higgs fields have masses
$ \sim 100 {\rm GeV} - 1 {\rm TeV} $.
\end{abstract}


\section{Introduction}
\label{sec:intro}

Extensions of the standard model may be motivated
in various points of view toward the discovery of new physics.
Among many intriguing possibilities,
the presence of isosinglet quarks is suggested
in certain models such as $ {\rm E}_6 $ type unified models
\cite{E6}.
Specifically, there are two types of singlet quarks,
$ U $ with electric charge $ Q_{\rm em} = 2/3 $
and $ D $ with $ Q_{\rm em} = - 1/3 $,
which may mix with the ordinary quarks.
Then, various novel features arise through the mixing between
the ordinary quarks ($ q = u , d $)
and the singlet quarks ($ Q = U , D $).
The unitarity of Cabibbo-Kobayashi-Maskawa (CKM) matrix
within the ordinary quark sector is violated,
and the flavor changing neutral currents (FCNC's) appear
at the tree-level
\cite{AgBr,LL,BB,qQmix,KY,Fr}.
These flavor changing interactions are actually described
in terms of the $ q $-$ Q $ mixing parameters and the quark masses,
as seen in detail in the text.
This may be viewed as an interesting extension
of the natural flavor conservation proposed in the early literature
\cite{GWP}.
Furthermore, the $ q $-$ Q $ mixing may involve $ CP $ violating phases.
Hence, it is quite expected that the $ q $-$ Q $ mixing
provide significant effects on various physical processes.
It is also noted that the so-called seesaw mechanism
even works for generating the ordinary quark masses
through the $ q $-$ Q $ mixing
\cite{qseesaw,BM,ssinverted}.

The $ q $-$ Q $ mixing effects on the $ Z $ boson mediated neutral currents
have been investigated so far extensively in the literature
\cite{LL,qQmix}.
These analyses show, in particular, that there is a good chance
to find the singlet quark effects in the $ B $ physics.
Some contributions of the neutral couplings mediated
by the Higgs scalar particles have also been considered
on the neutron electric dipole moment and the neutral meson mixings
\cite{BB,KY,BM}.

The singlet quarks may even provide important contributions
in cosmology.  In fact, for the electroweak baryogenesis
\cite{baryogenesis}
the $ CP $ violating $ q $-$ Q $ mixing through the coupling
with a complex singlet Higgs field $ S $ can be efficient
to generate the chiral charge fluxes through the bubble wall
\cite{tUbaryogenesis,bDbaryogenesis}.
This possibility is encouraging,
since the $ CP $ asymmetry induced by the conventional CKM phase
is far too small to account for the observed baryon to entropy ratio.
Furthermore, the singlet Higgs field $ S $
providing the singlet quark mass term and the $ q $-$ Q $ mixing term
is preferable for realizing
a strong enough first order electroweak phase transition
\cite{bDbaryogenesis}.

As mentioned so far, the singlet quarks may bring
various intriguing features in particle physics and cosmology.
Then, it is worth understanding in detail
the characteristic properties of the electroweak models
incorporating the singlet quarks.
Specifically, it is important to show
how the ordinary quark masses and mixings are affected
by the $ q $-$ Q $ mixing.
The structures of the CKM mixing and FCNC's
should also be clarified properly.
In this article, we present systematic and comprehensive descriptions
of the quark masses, mixings and flavor changing interactions
in the presence of singlet quarks.
In Sec. \ref{sec:model}, a representative model
with singlet quarks is presented.
In Sec. \ref {sec:mass-mixing}, the quark masses and mixings,
which are affected by the $ q $-$ Q $ mixing, are calculated in detail.
Then, the $ q $-$ Q $ mixing effects on the gauge and scalar interactions
are examined in Sec. \ref{sec:FCI}.
They are described appropriately
in terms of the $ q $-$ Q $ mixing parameters and the quark masses.
In Sec. \ref{sec:numerical}, numerical calculations are performed
to confirm the flavor structures of the $ q $-$ Q $ mixing effects.
Sec. \ref{sec:summary} is devoted to the summary
and some discussion on the phenomenological effects
provided by the singlet quarks.
The technical details in diagonalizing the quark mass matrix
are presented in Appendix \ref{sec:diagonalization}.

\section{Electroweak model with singlet quarks}
\label{sec:model}

We first describe a representative electroweak model
based on the gauge symmetry
$ {\rm SU(3)}_C \times {\rm SU(2)}_W \times {\rm U(1)}_Y $,
where both types of singlet quarks $ U $ and $ D $
and also one complex singlet Higgs field $ S $ are incorporated.
The generic Yukawa couplings of quarks are given by
\begin{eqnarray}
{\cal L}_{\rm Y}
&=& - \ u^c_0 \lambda_u \Psi_{q_0} \Phi_H - U^c_0 h_u \Psi_{q_0} \Phi_H
\nonumber \\
& & - \ u^c_0 ( f_U S + f_U^\prime S^\dagger ) U_0
- U^c_0 ( \lambda_U S + \lambda_U^\prime S^\dagger ) U_0
\nonumber \\
& & - \ d^c_0 \lambda_d V_0^\dagger \Psi_{q_0} {\tilde \Phi_H}
- D^c_0 h_d V_0^\dagger \Psi_{q_0} {\tilde \Phi_H}
\nonumber \\
& & - \ d^c_0 ( f_D S + f_D^\prime S^\dagger ) D_0
- D^c_0 ( \lambda_D S + \lambda_D^\prime S^\dagger ) D_0
\nonumber \\
& & + \ {\rm h.c.}
\label{eqn:LYukawa}
\end{eqnarray}
in terms of the two-component Weyl fields
for the weak eigenstates with subscript ``0".
(The generation indices and the Lorentz factors are omitted
here for simplicity.)
The isodoublets of left-handed ordinary quarks are represented by
\begin{equation}
\Psi_{q_0}
= \left( \begin{array}{c} u_0 \\ V_0 d_0 \end{array} \right)
\end{equation}
with a certain $ 3 \times 3 $ unitary matrix $ V_0 $, and
$ V_0^\dagger \Psi_{q_0} \equiv ( V_0^\dagger u_0 , d_0 )^{\rm T} $.
The Higgs doublet is also given by
\begin{equation}
\Phi_H = \left( \begin{array}{c} H^+ \\
H^0 \end{array} \right)
\end{equation}
with $ {\tilde \Phi}_H \equiv i \tau_2 \Phi_H^* $.

The Higgs fields develop vacuum expectation values (VEV's),
\begin{equation}
\langle H^0 \rangle = \frac{v}{\sqrt 2} , \
\langle S \rangle = \frac{v_S}{\sqrt 2} \ {\rm e}^{i \phi_S} .
\end{equation}
Here a complex phase $ \phi_S $ is included in $ \langle S \rangle $,
which may be induced by the $ CP $ violation in the Higgs sector
either spontaneous or explicit.  The quark mass matrix is then produced
with these VEV's as
\begin{equation}
{\cal M}_{\cal Q} = \left( \begin{array}{cc}
M_q & \Delta_{qQ} \\
   \Delta_{qQ}^\prime & M_Q \end{array} \right) ,
\end{equation}
where the submatrices are given by
\begin{eqnarray}
M_q &=& \frac{1}{\sqrt 2} \lambda_q v ,
\label{eqn:Mq} \\
\Delta_{qQ} &=& \frac{1}{\sqrt 2}
( f_Q {\rm e}^{i \phi_S} + f_Q^\prime {\rm e}^{-i \phi_S} ) v_S ,
\label{eqn:DqQ} \\
\Delta_{qQ}^\prime &=& \frac{1}{\sqrt 2} h_q v ,
\label{eqn:DqQp} \\
M_Q &=& \frac{1}{\sqrt 2}
( \lambda_Q {\rm e}^{i \phi_S}
+ \lambda_Q^\prime {\rm e}^{-i \phi_S} ) v_S .
\label{eqn:MQ}
\end{eqnarray}
Hereafter, the quarks with the same electric charge are collectively
denoted by $ {\cal Q} = ( q , Q ) $, i.e.,
$ {\cal U} = ( u , U ) $ and $ {\cal D} = (d,D) $.
The dimensions of $ {\cal M}_{\cal Q} $ and its submatrices
are specified with the three generations of ordinary quarks
and the number $ N_Q $ of singlet quarks.

Some remarks may be presented concerning possible variants of the model.
It is straightforward to describe the models
admitting only either $ U $ or $ D $ quarks.
The complex singlet Higgs field $ S $
is employed in the present model to generate the singlet quark mass term
and $ q $-$ Q $ mixing term.
This choice will be motivated, in particular, for the electroweak
baryogenesis.  The $ CP $ violating $ q $-$ Q $ mixing
provided by the space-dependent complex Higgs field $ S $
can be efficient for generating the chiral charge fluxes
through the bubble wall
\cite{tUbaryogenesis,bDbaryogenesis}.
Alternative options, however, may be considered
about the singlet Higgs $ S $, including the following cases.
(i) The singlet Higgs $ S $ is a real scalar field rather than complex one.
(ii) The singlet Higgs $ S $ is eliminated
(or it has a mass much larger than the electroweak scale)
while leaving its contributions to the quark mass matrix.
(iii) The models with the complex Higgs $ S $
may be extended by incorporating supersymmetry.
The following investigations are made in detail
for the model given in Eq. (\ref{eqn:LYukawa}) with one complex $ S $.
The results can be extended readily for these variants of the model,
as will be mentioned occasionally.

\section{Quark masses and mixings with singlet quarks}
\label{sec:mass-mixing}

We present in this section the detailed description of the quark masses
and mixings which are affected by the $ q $-$ Q $ mixing.
It is desirable that even in the presence of singlet quarks
the ordinary quark mass hierarchy and the CKM structure
are reproduced in some reasonable regions of the model parameter space.
This issue will be addressed in the following by inspecting systematically
the form of quark mass matrix $ {\cal M}_{\cal Q} $
and its diagonalization.

The total quark mass matrix is diagonalized as usual
by unitary transformations $ {\cal V}_{{\cal Q}_{\rm L}} $
and $ {\cal V}_{{\cal Q}_{\rm R}} $:
\begin{equation}
{\cal V}_{{\cal Q}_{\rm R}}^\dagger
{\cal M}_{\cal Q} {\cal V}_{{\cal Q}_{\rm L}}
= \left( \begin{array}{cc}
{\bar M}_q & {\bf 0} \\ {\bf 0} & {\bar M}_Q
\end{array} \right) ,
\label{eqn:MQdiagonal}
\end{equation}
where
\begin{eqnarray}
{\bar M}_q &=& {\rm diag.}( m_{q_1} , m_{q_2} , m_{q_3} ) ,
\\
{\bar M}_Q &=& {\rm diag.}( m_{Q_1} , \cdots ) ,
\end{eqnarray}
and $ ( q_1 , q_2 , q_3 ) = ( u , c , t ) $ or $ ( d , s , b ) $.
The quark mass eigenstates are given by
\begin{eqnarray}
\left( \begin{array}{c} q \\ Q \end{array} \right)
&=& {\cal V}_{{\cal Q}_{\rm L}}^\dagger
\left( \begin{array}{c} q_0 \\ Q_0 \end{array} \right) ,
\\
( q^c , Q^c ) &=& ( q_0^c , Q_0^c ){\cal V}_{{\cal Q}_{\rm R}} .
\end{eqnarray}
The relevant $ ( 3 + N_Q ) \times ( 3 + N_Q ) $ unitary matrices
are given by
\begin{equation}
{\cal V}_{{\cal Q}_\chi}
= \left( \begin{array}{cc}
V_{q_\chi} & \epsilon_{q_\chi} \\
- \epsilon_{q_\chi}^{\prime \dagger} & V_{Q_\chi}
\end{array} \right)
\label{eqn:VQchi}
\end{equation}
for the respective chirality sectors $ \chi = {\rm L} , {\rm R} $.
The $ 3 \times N_Q $ matrices $ \epsilon_{q_\chi} $
and $ \epsilon_{q_\chi}^\prime $ represent the $ q $-$ Q $ mixing.

The leading order calculations on the quark masses and mixings
with singlet quarks are given for $ N_Q = 1 $ in the literature
\cite{AgBr}.
Here, we would rather like to present comprehensive understandings
on the $ q $-$ Q $ mixing effects which are even valid
beyond the leading orders for more general cases
including several singlet quarks ($ N_Q \geq 1 $).

\subsection{Choices of the quark basis}

We first note that the quark mass matrix $ {\cal M}_{\cal Q} $
may be reduced to a specific form with either $ \Delta_{qQ} = {\bf 0} $
or $ \Delta_{qQ}^\prime = {\bf 0} $ by a unitary transformation
of the right-handed quarks, which are undistinguishable
by means of the $ {\rm SU(3)}_C \times {\rm SU(2)}_W \times {\rm U(1)}_Y $.
Then, the Yukawa coupling $ \lambda_q $ can be made
diagonal and non-negative by unitary transformations of the ordinary quarks:
\begin{equation}
\lambda_q = {\rm diag.}( \lambda_{q_1} , \lambda_{q_2} , \lambda_{q_3} ) .
\end{equation}
The other couplings $ f_Q $, $ f_Q^\prime $ and $ h_q $
involving the ordinary quarks
as well as the mixing matrix $ V_0 $ are redefined accordingly.
The condition $ \Delta_{qQ} = {\bf 0} $ or $ \Delta_{qQ}^\prime = {\bf 0} $
is, however, maintained by these transformations of the ordinary quarks.
In this basis, by turning off the $ q $-$ Q $ mixing
with $ f_Q , f_Q^\prime , h_q \rightarrow 0 $,
the quark fields $ u_0 $ and $ d_0 $ are reduced to the mass eigenstates,
and $ V_0 $ is identified with the CKM matrix.
The actual CKM matrix $ V $ is slightly modified from $ V_0 $
by the $ q $-$ Q $ mixing, as shown explicitly later.

It should be noticed that the quark transformations made so far
to specify the form of $ {\cal M}_{\cal Q} $
do not mix the electroweak doublets with the singlets,
respecting the $ {\rm SU(3)}_C \times {\rm SU(2)}_W \times {\rm U(1)}_Y $
gauge symmetry.  Hence, without loss of generality
we may have two appropriate choices of the quark basis
for the electroweak eigenstates:
\begin{eqnarray}
{\mbox{\bf basis (a)}} & : & \Delta^\prime_{qQ} = {\bf 0} ,
\\
{\mbox{\bf basis (b)}} & : & \Delta_{qQ} = {\bf 0} .
\end{eqnarray}
In either case, with the diagonal $ \lambda_q $ coupling
we have the submatrix of $ {\cal M}_{\cal Q} $,
\begin{equation}
M_q = {\rm diag.}( m_{q_1}^0 , m_{q_2}^0 , m_{q_3}^0 ) ,
\end{equation}
with
\begin{equation}
m_{q_i}^0 = \lambda_{q_i} v / {\sqrt 2} .
\end{equation}

As seen from Eq. (\ref{eqn:DqQp}), the basis (a) is chosen naturally
by eliminating the $ h_q $ coupling with a unitary transformation
of the right-handed quarks:
\begin{equation}
h_q = {\bf 0} \rightarrow \Delta_{qQ}^\prime = {\bf 0} .
\label{eqn:hq0}
\end{equation}
This choice is in fact made
irrespectively of the specific values of the VEV's.
On the other hand, the condition,
\begin{equation}
f_Q {\rm e}^{i \phi_S} + f_Q^\prime {\rm e}^{-i \phi_S} = {\bf 0}
\rightarrow \Delta_{qQ} = {\bf 0} ,
\label{eqn:fQ0}
\end{equation}
for the basis (b) seems to require some tuning
between $ f_Q $ and $ f_Q^\prime $, which depends on the phase $ \phi_S $
of $ \langle S \rangle $.
This tuning may, however, be evaded in some cases
including one real $ S $, no $ S $ and one supersymmetric $ S $ models.
In these models, the $ f_Q^\prime $ coupling is absent,
and then the $ \Delta_{qQ} $ term is rotated out naturally
together with the $ f_Q $ coupling.

The condition $ M_q = {\bf 0} $ may even be realized
by means of the symmetries and matter contents
so as to distinguish the ordinary quarks from the singlet quarks.
This is in fact the case in some left-right gauge models.
Then, the so-called seesaw mechanism is available
for generating the ordinary quark masses with three singlet quarks
\cite{qseesaw,BM,ssinverted}:
\begin{equation}
{\mbox{\bf seesaw}} \ : \ M_q = {\bf 0} \ ( N_Q = 3 ) .
\end{equation}
The seesaw case may formally be reduced to the basis (a)
by the exchange $ q^c \leftrightarrow Q^c $ of the right-handed quarks.
It will, however, be appropriate to treat separately the seesaw case
in its own right.

In the following, we examine accurately for the respective bases
how the quark masses and mixings are affected by the $ q $-$ Q $ mixing.
The technical details in diagonalizing $ {\cal M}_{\cal Q} $
are presented in Appendix \ref{sec:diagonalization}.

\subsection{Basis (a) with $ \Delta_{qQ}^\prime = {\bf 0} $}

In the basis (a) with $ \Delta_{qQ}^\prime = {\bf 0} $,
the quark mass matrix is given by
\begin{equation}
{\cal M}_{\cal Q} = \left( \begin{array}{cc}
M_q & \Delta_{qQ} \\ {\bf 0} & M_Q \end{array} \right) .
\end{equation}
Then, it is relevant to introduce the $ q $-$ Q $ mixing parameters
\begin{eqnarray}
\epsilon^f_i &=& {\rm max} \left[ | ( \Delta_{qQ} )_{ia} | / m_Q \right]
\nonumber \\
&\sim& ( v_S / m_Q )
( | ( f_Q )_{ia} | + | ( f_Q^\prime )_{ia} | ) ,
\end{eqnarray}
each of which represents the magnitude of the mixing
between the $ i $-th ordinary quark and the singlet quarks.
Here, it is assumed that the singlet quarks have masses
$ \sim m_Q $ of the same order.

The ordinary quark masses are obtained as
\begin{equation}
m_{q_i} = c_i ( {\mbox{\boldmath $ \epsilon $}}^f ) \lambda_{q_i} v ,
\label{eqn:mq-a}
\end{equation}
where $ c_i ( {\mbox{\boldmath $ \epsilon $}}^f ) \sim 1 $
depending on $ {\mbox{\boldmath $ \epsilon $}}^f
\equiv ( \epsilon^f_1 , \epsilon^f_2 , \epsilon^f_3 ) $.
(See Appendix \ref{sec:diagonalization} for the detailed arguments.) 
The $ q $-$ Q $ mixing and the ordinary quark mixing
are estimated in magnitude as
\begin{eqnarray}
( \epsilon_{q_{\rm L}} )_{ia} & \sim & ( \epsilon_{q_{\rm L}}^\prime )_{ia}
\sim ( m_{q_i} / m_Q ) \epsilon^f_i ,
\label{eqn:eqL-a} \\
( \epsilon_{q_{\rm R}} )_{ia} & \sim & ( \epsilon_{q_{\rm R}}^\prime )_{ia}
\sim \epsilon^f_i ,
\label{eqn:epR-a} \\
( V_{q_{\rm L}} )_{ij} & \sim & \delta_{ij}
+ \frac{m_{q_i} m_{q_j}}{m_{q_i}^2 + m_{q_j}^2}
\epsilon^f_i \epsilon^f_j ,
\label{eqn:VqL-a} \\
( V_{q_{\rm R}} )_{ij} & \sim & \delta_{ij}
+ \frac{m_{q_j}^2}{m_{q_i}^2 + m_{q_j}^2}
\epsilon^f_i \epsilon^f_j ,
\label{eqn:VqR-a}
\end{eqnarray}
where
\begin{eqnarray}
\frac{m_{q_i} m_{q_j}}{m_{q_i}^2 + m_{q_j}^2}
& \approx & \left\{ \begin{array}{ll}
m_{q_i} / m_{q_j} & ( m_{q_i} \ll m_{q_j} ) \\
{ } & { } \\
m_{q_j} / m_{q_i} & ( m_{q_i} \gg m_{q_j} )
\end{array} \right. ,
\\
\frac{m_{q_j}^2}{m_{q_i}^2 + m_{q_j}^2}
& \approx & \left\{ \begin{array}{ll}
1 & ( m_{q_i} \ll m_{q_j} ) \\
{ } & { } \\
( m_{q_j} / m_{q_i} )^2 & ( m_{q_i} \gg m_{q_j} )
\end{array} \right. .
\end{eqnarray}
The corrections to the diagonal components of $ V_{q_\chi} $
are estimated precisely in terms of the $ q $-$ Q $ mixing matrices
by noting the unitarity relations for the whole transformation matrices
$ {\cal V}_{{\cal Q}_\chi} $ given in Eq. (\ref{eqn:VQchi}):
\begin{eqnarray}
V_{q_\chi}^\dagger V_{q_\chi}
&=& {\bf 1} - \epsilon_{q_\chi}^\prime \epsilon_{q_\chi}^{\prime \dagger} ,
\label{eqn:unvVq1} \\
V_{q_\chi} V_{q_\chi}^\dagger
&=& {\bf 1} - \epsilon_{q_\chi} \epsilon_{q_\chi}^\dagger .
\label{eqn:unvVq2}
\end{eqnarray}

We have really observed in Eq. (\ref{eqn:mq-a}) that
the hierarchical masses $ m_{q_i} $ of the ordinary quarks
are reproduced by the corresponding Yukawa couplings
$ \lambda_{q_i} $ even in the presence of the $ q $-$ Q $ mixing.
The $ q $-$ Q $ mixing effects
are described in terms of the parameters $ \epsilon^f_i $
in the basis (a) with $ \Delta_{qQ}^\prime = {\bf 0} $.
It should particularly be noted in Eq. (\ref{eqn:eqL-a})
that the left-handed $ q $-$ Q $ mixing
is suppressed further by the $ q/Q $ mass ratios $ m_{q_i} / m_Q $.
The ordinary quark mixings $ ( V_{q_\chi} )_{ij} $ ($ i \not= j $)
arising at the order of $ \epsilon^f_i \epsilon^f_j $
are actually related to the ordinary quark mass ratios
in Eqs. (\ref{eqn:VqL-a}) and (\ref{eqn:VqR-a}).
The unitarity violation of $ V_{q_\chi} $
is determined by the $ q $-$ Q $ mixing matrices
$ \epsilon_{q_\chi} $ and $ \epsilon_{q_\chi}^\prime $
in Eqs. (\ref{eqn:unvVq1}) and (\ref{eqn:unvVq2}).
These features of the $ q $-$ Q $ mixing effects,
as seen more explicitly in the leading order calculations
\cite{AgBr}, even hold for $ N_Q \geq 1 $ beyond the leading orders.

\subsection{Basis (b) with $ \Delta_{qQ} = {\bf 0} $}

In the basis (b) with $ \Delta_{qQ} = {\bf 0} $,
the quark mass matrix is given by
\begin{equation}
{\cal M}_{\cal Q} = \left( \begin{array}{cc}
M_q & {\bf 0} \\ \Delta_{qQ}^\prime & M_Q \end{array} \right) .
\end{equation}
Then, the $ q $-$ Q $ mixing parameters are taken as
\begin{eqnarray}
\epsilon^h_i
&=& {\rm max} \left[ | ( \Delta_{qQ}^\prime )_{ai} | / m_Q \right]
\nonumber \\
& \sim & ( v / m_Q ) | ( h_q )_{ai} | .
\end{eqnarray}
The ordinary quark masses are obtained as
\begin{equation}
m_{q_i} = c^\prime_i ( {\mbox{\boldmath $ \epsilon $}}^h ) \lambda_{q_i} v ,
\label{eqn:mq-b}
\end{equation}
where $ c^\prime_i ( {\mbox{\boldmath $ \epsilon $}}^h ) \sim 1 $
depending on $ {\mbox{\boldmath $ \epsilon $}}^h
\equiv ( \epsilon^h_1 , \epsilon^h_2 , \epsilon^h_3 ) $.
The $ q $-$ Q $ mixing effects are estimated as
\begin{eqnarray}
( \epsilon_{q_{\rm L}} )_{ia} & \sim & ( \epsilon_{q_{\rm L}}^\prime )_{ia}
\sim \epsilon^h_i ,
\label{eqn:eqL-b} \\
( \epsilon_{q_{\rm R}} )_{ia} & \sim & ( \epsilon_{q_{\rm R}}^\prime )_{ia}
\sim ( m_{q_i} / m_Q ) \epsilon^h_i ,
\label{eqn:eqR-b} \\
( V_{q_{\rm L}} )_{ij} & \sim & \delta_{ij}
+ \frac{m_{q_j}^2}{m_{q_i}^2 + m_{q_j}^2}
\epsilon^h_i \epsilon^h_j ,
\label{eqn:VqL-b} \\
( V_{q_{\rm R}} )_{ij} & \sim & \delta_{ij}
+ \frac{m_{q_i} m_{q_j}}{m_{q_i}^2 + m_{q_j}^2}
\epsilon^h_i \epsilon^h_j .
\label{eqn:VqR-b}
\end{eqnarray}
Here, the left-handed $ q $-$ Q $ mixing in Eq. (\ref{eqn:eqL-b})
is no longer suppressed by the $ q/Q $ mass ratios
in contrast to Eq. (\ref{eqn:eqL-a}) for the basis (a).
Then, as described explicitly in the next section,
the CKM unitarity violation and the $ Z $ mediated FCNC's
can be significant for reasonable values
of the $ q $-$ Q $ mixing parameters $ \epsilon^h_i $
\cite{LL,qQmix}.

We find again in Eq. (\ref{eqn:mq-b}) that the actual masses
of the ordinary quarks are reproduced
in terms of the relevant Yukawa couplings.
It should, however, be mentioned that this relation
might not be stable generally.
In fact, the quark basis (b) is found by eliminating the $ q^c Q $ term
$ \Delta_{qQ} = f_Q \langle S \rangle + f_Q^\prime \langle S \rangle^* $.
This seems to be a fine tuning between the couplings
$ f_Q $ and $ f_Q^\prime $ with a given $ \langle S \rangle $.
If the complex phase of $ \langle S \rangle $ is changed
by radiative corrections, the quark basis (b) is rearranged
so as to keep the condition $ \Delta_{qQ} = {\bf 0} $.
Accordingly, the new diagonal coupling $ {\tilde \lambda}_q $
is obtained as a mixture of $ \lambda_q $ and $ h_q $.
Then, even if the original couplings $ \lambda_{q_i} $
are chosen so as to reproduce the hierarchical quark masses
$ m_{q_i} \sim \lambda_{q_i} v $,
there is in general no guarantee that the new couplings
$ {\tilde \lambda}_{q_i} $ also take the similar values.
This problem encountered in the basis (b) may, however,
be evaded reasonably in certain cases.
For example, in the models without the $ f_Q^\prime $ coupling,
the $ f_Q $ coupling is eliminated from the beginning
by using the right-handed quark transformation.
It is also considered that the $ \lambda_q $ and $ h_q $ couplings
have the same flavor structure as
$ ( h_q )_{ai} \sim \lambda_{q_i} \sim m_{q_i} / v $.
This relation between $ \lambda_q $ and $ h_q $
is in fact technically natural, and the new $ {\tilde \lambda}_q $
also has the same hierarchical flavor structure.

\subsection{Seesaw model}

The quark mass matrix of the seesaw form ($ N_Q = 3 $)
\cite{qseesaw,BM,ssinverted} is given by
\begin{equation}
{\cal M}_{\cal Q} = \left( \begin{array}{cc}
{\bf 0} & \Delta_{qQ} \\ \Delta_{qQ}^\prime & M_Q \end{array} \right) .
\end{equation}
We first consider the case where the singlet quarks have comparable masses,
\begin{equation}
m_{Q_1} , \ m_{Q_2} , \ m_{Q_3} \sim m_Q .
\end{equation}
Then, the ordinary quark masses are given by
\begin{equation}
m_{q_i} = c^{\prime \prime}_i
( {\mbox{\boldmath $ \epsilon $}}^f , {\mbox{\boldmath $ \epsilon $}}^h )
\epsilon^f_i \epsilon^h_i m_Q ,
\label{eqn:mq-sss}
\end{equation}
where $ c^{\prime \prime}_i
( {\mbox{\boldmath $ \epsilon $}}^f , {\mbox{\boldmath $ \epsilon $}}^h )
\sim 1 $.  The $ q $-$ Q $ mixing effects are estimated as
\begin{eqnarray}
( \epsilon_{q_{\rm L}} )_{ia} & \sim & ( \epsilon_{q_{\rm L}}^\prime )_{ia}
\sim \epsilon^h_i ,
\label{eqn:eqL-sss} \\
( \epsilon_{q_{\rm R}} )_{ia} & \sim & ( \epsilon_{q_{\rm R}}^\prime )_{ia}
\sim \epsilon^f_i ,
\label{eqn:eqR-sss} \\
( V_{q_{\rm L}} )_{ij} & \sim & \delta_{ij}
+ \frac{\epsilon^h_i \epsilon^h_j}{( \epsilon^h_i )^2 + ( \epsilon^h_j )^2} ,
\label{eqn:VqL-sss} \\
( V_{q_{\rm R}} )_{ij} & \sim & \delta_{ij}
+ \frac{\epsilon^f_i \epsilon^f_j}{( \epsilon^f_i )^2 + ( \epsilon^f_j )^2} .
\label{eqn:VqR-sss}
\end{eqnarray}

The actual masses of the ordinary quarks may be reproduced
in Eq. (\ref{eqn:mq-sss}) under the hierarchy
for the $ q $-$ Q $ mixing terms,
\begin{equation}
| ( \Delta_{qQ} )_{1a} | \ll | ( \Delta_{qQ} )_{2b} | \ll
| ( \Delta_{qQ} )_{3c} |
\label{eqn:DqQ123}
\end{equation}
and/or
\begin{equation}
| ( \Delta_{qQ}^\prime )_{a1} | \ll | ( \Delta_{qQ}^\prime )_{b2} | \ll
| ( \Delta_{qQ}^\prime )_{c3} | .
\label{eqn:DqQp123}
\end{equation}
Then, by considering some typical cases,
the quark mixings provided by the singlet quarks
may be related to the quark masses with Eq. (\ref{eqn:mq-sss}):
\begin{eqnarray}
{\mbox{(i)}}
&:& \epsilon^f_i \sim \epsilon^h_i \sim {\sqrt{ m_{q_i} / m_Q }} ,
\nonumber \\
& & ( V_{q_{\rm L}} )_{ij} \sim ( V_{q_{\rm R}} )_{ij} \sim
\frac{\sqrt{ m_{q_i} m_{q_j} }}{ m_{q_i} + m_{q_j} } .
\label{eqn:qQmix-i}
\\
{\mbox{(ii)}} &:& \epsilon^f_i \sim m_{q_i} / {\bar h} v , \
\epsilon^h_i \sim {\bar h} v / m_Q ,
\nonumber \\
& & ( V_{q_{\rm L}} )_{ij} \sim 1 , \
( V_{q_{\rm R}} )_{ij} \sim
\frac{m_{q_i} m_{q_j}}{m_{q_i}^2 + m_{q_j}^2} .
\label{eqn:qQmix-ii}
\\
{\mbox{(iii)}} &:& \epsilon^f_i \sim {\bar f} v_S / m_Q , \
\epsilon^h_i \sim m_{q_i} / {\bar f} v_S ,
\nonumber \\
& & ( V_{q_{\rm L}} )_{ij} \sim
\frac{m_{q_i} m_{q_j}}{m_{q_i}^2 + m_{q_j}^2} , \
( V_{q_{\rm R}} )_{ij} \sim 1 .
\label{eqn:qQmix-iii}
\end{eqnarray}
In the case (i) the $ q $-$ Q $ mixing terms
$ \Delta_{qQ} $ and $ \Delta_{qQ}^\prime $ are assumed
to have the same flavor structure.
In the case (ii) only the $ \Delta_{qQ} $ term
has the hierarchical form (\ref{eqn:DqQ123})
while the $ \Delta_{qQ}^\prime $ term $ \sim {\bar h} v $
is rather flavor-independent.
On the other hand, in the case (iii) only the $ \Delta_{qQ}^\prime $ term
has the hierarchical form (\ref{eqn:DqQp123})
while the $ \Delta_{qQ} $ term $ \sim {\bar f} v_S $
is rather flavor-independent.

While the above choice for the model parameters is technically natural,
there is another attractive possibility for realizing
the ordinary quark mass hierarchy \cite{ssinverted}.
That is, the inverted hierarchy is assumed for the singlet quark masses,
\begin{equation}
m_{Q_1} \gg m_{Q_2} \gg m_{Q_3} .
\label{eqn:mQ-ssi}
\end{equation}
The specific relations such as Eqs. (\ref{eqn:DqQ123})
and (\ref{eqn:DqQp123}) are, however, not invoked
on the $ q $-$ Q $ mixing terms, i.e.,
\begin{equation}
( \Delta_{qQ} )_{ia} \sim {\bar f} v_S , \
( \Delta_{qQ}^\prime )_{ai} \sim {\bar h} v ,
\end{equation}
where $ {\bar f} $ and $ {\bar h} $ represent
the mean values of the relevant Yukawa couplings.
In this situation, the ordinary quark masses are given by
\begin{equation}
m_{q_i} = c^{\prime \prime \prime}_i ( {\bar f} , {\bar h} )
( {\bar f} v_S / m_{Q_i} ) {\bar h} v ,
\label{eqn:mq-ssi}
\end{equation}
where $ c^{\prime \prime \prime}_i ( {\bar f} , {\bar h} ) \sim 1 $
\cite{ssinverted}.
Accordingly, the quark mixings are obtained
in terms of the ordinary quark masses as
\begin{eqnarray}
( {\tilde V}_{q_{\rm L}}^\dagger \epsilon_{q_{\rm L}} )_{ia}
& \sim & ( \epsilon_{q_{\rm L}}^\prime )_{ia}
\sim ( m_{q_i} / {\bar f} v_S ) ,
\label{eqn:eqL-ssi} \\
( {\tilde V}_{q_{\rm R}}^\dagger \epsilon_{q_{\rm R}} )_{ia}
& \sim & ( \epsilon_{q_{\rm R}}^\prime )_{ia}
\sim ( m_{q_i} / {\bar h} v ) ,
\label{eqn:eqR-ssi} \\
( {\tilde V}_{q_{\rm L}}^\dagger V_{q_{\rm L}} )_{ij} & \sim &
\frac{m_{q_i} m_{q_j}}{m_{q_i}^2 + m_{q_j}^2} ,
\label{eqn:VqL-ssi} \\
( {\tilde V}_{q_{\rm R}}^\dagger V_{q_{\rm R}} )_{ij} & \sim &
\frac{m_{q_i} m_{q_j}}{m_{q_i}^2 + m_{q_j}^2} .
\label{eqn:VqR-ssi}
\end{eqnarray}
Here, the unitary matrices $ {\tilde V}_{q_{\rm L}}^\dagger $
and $ {\tilde V}_{q_{\rm R}}^\dagger $ are introduced to deform
the $ \Delta_{qQ} $ and $ \Delta_{qQ}^\prime $ terms
to the triangular forms as given
in Eqs. (\ref{eqn:DqQ-ssi}) and (\ref{eqn:DqQp-ssi}).

It is indeed seen in Eq. (\ref{eqn:mq-ssi})
that the ordinary quark mass hierarchy is realized
by the inverted hierarchy (\ref{eqn:mQ-ssi}) of singlet quarks.
These relations for the quark masses and mixings
are essentially applicable even for the case $ m_T^0 \ll {\bar f} v_S $
($ m_T^0 $ is the third component of the diagonal $ M_Q $),
as mentioned in Ref. \cite{ssinverted}.
In this case, we obtain
\begin{equation}
m_t \sim {\bar h} v , \ m_T \sim {\bar f} v_S
\end{equation}
with the significant right-handed $ t $-$ T $ mixing
with $ ( \epsilon_{q_{\rm R}} )_{tT} , ( \epsilon_{q_{\rm R}}^\prime )_{tT}
\simeq 1 $.

\subsection{Relations among the three bases}

We have described so far the structures of the $ q $-$ Q $ mixing effects
on the quark masses and mixings.  They are summarized
in Table \ref{table:mixing-mass}. 
We here discuss the relations among these three representative bases.

The basis (a) with $ \Delta_{qQ}^\prime = {\bf 0} $
is suitable as long as the $ \Delta_{qQ} $ term does not
exceed the $ M_Q $ term.
In this case, the quark masses and mixings are described well
in terms of $ \epsilon^f_i \sim | ( \Delta_{qQ} )_{ia} | / m_Q \lesssim 1 $.
This treatment is, however, invalidated
when $ | \Delta_{qQ} | \gg | M_Q | $ in the basis (a).
Then, we may seek a more appropriate quark basis
by making some unitary transformations of the quarks.
These transformations should respect
the $ {\rm SU(3)}_C \times {\rm SU(2)}_W \times {\rm U(1)}_Y $, i.e.,
the components of the doublets are not mixed with the singlets
in the left-handed quark sector
before the quark mass terms are generated by the spontaneous breakdown
of the $ {\rm SU(2)}_W \times {\rm U(1)}_Y $.

In general, for any $ N_Q \geq 1 $ the quark mass matrix in the basis (a)
can be deformed as follows by a unitary transformation
$ {\tilde {\cal V}}_{{\cal Q}_{\rm R}}^{(1)} $ of the right-handed quarks:
\begin{equation}
{\tilde {\cal V}}_{{\cal Q}_{\rm R}}^{(1) \dagger}
\left( \begin{array}{cc}
M_q & \Delta_{qQ} \\ {\bf 0} & M_Q \end{array} \right)
= \left( \begin{array}{cc} {\tilde M}_q^{(1)} & {\bf 0} \\
{\tilde \Delta}_{qQ}^{(1) \prime} & {\tilde M}_Q^{(1)} \end{array} \right) .
\end{equation}
The $ q $-$ Q $ mixing in $ {\tilde {\cal V}}_{{\cal Q}_{\rm R}}^{(1)} $
is significant with $ {\tilde \epsilon}_{q_{\rm R}}^{(1)} \sim 1 $
for the case of $ | \Delta_{qQ} | \gtrsim | M_Q | $.
Here, we note the relation between $ {\tilde M}_q^{(1)} $
and $ {\tilde \Delta}_{qQ}^{(1) \prime} $ both of which are
provided by the original $ M_q $:
\begin{equation}
{\tilde \Delta}_{qQ}^{(1) \prime}
= {\tilde \epsilon}_{q_{\rm R}}^{(1) \dagger} 
{\tilde V}_{q_{\rm R}}^{(1) \dagger -1} {\tilde M}_q^{(1)} .
\end{equation}
The mass matrix $ {\tilde M}_q^{(1)} $ is then diagonalized
by the unitary transformations $ {\tilde V}_{q_{\rm L}}^{(2)} $
and $ {\tilde V}_{q_{\rm R}}^{(2)} $ of the ordinary quarks as
\begin{eqnarray}
{\tilde M}_q &=& {\tilde V}_{q_{\rm R}}^{(2) \dagger}
{\tilde M}_q^{(1)} {\tilde V}_{q_{\rm L}}^{(2)}
\nonumber \\
&=& {\rm diag.}
( {\tilde m}_{q_1}^0 , {\tilde m}_{q_2}^0 , {\tilde m}_{q_3}^0 ) .
\end{eqnarray}
Accordingly, the $ {\tilde \Delta}_{qQ}^{(1) \prime} $ term
is deformed as
\begin{equation}
{\tilde \Delta}_{qQ}^\prime
= {\tilde \Delta}_{qQ}^{(1) \prime} {\tilde V}_{q_{\rm L}}^{(2)}
= {\tilde \epsilon}_{q_{\rm R}}^{(1) \dagger}
{\tilde V}_{q_{\rm R}}^{(1) \dagger -1} {\tilde V}_{q_{\rm R}}^{(2)}
{\tilde M}_q .
\label{eqn:DqQp-Mq}
\end{equation}
In this way, starting with the quark mass matrix
$ {\cal M}_{\cal Q} $ in the basis (a)
with $ \Delta_{qQ}^\prime = {\bf 0} $,
we have obtained the quark mass matrix
$ {\tilde {\cal M}}_{\cal Q} $ in the basis (b)
with $ {\tilde \Delta}_{qQ} = {\bf 0} $.

Here, it is noticed in Eq. (\ref{eqn:DqQp-Mq})
that the $ {\tilde \Delta}_{qQ}^\prime $ term
is related to the ordinary quark mass matrix $ {\tilde M}_q $ as
\begin{equation}
( {\tilde \Delta}_{qQ}^\prime )_{ai} \sim ( {\tilde M}_q )_{ii} .
\end{equation}
This implies the relation of the Yukawa couplings,
\begin{equation}
( {\tilde h}_q )_{ai} \sim {\tilde \lambda}_{q_i} .
\label{eqn:hq-lmq}
\end{equation}
Then, the quark masses are obtained as
$ m_{q_i} \sim {\tilde m}^0_{q_i} = ( {\tilde M}_q )_{ii} $,
and the $ q $-$ Q $ mixing parameters are specifically
given by the $ q/Q $ mass ratios as
\begin{equation}
{\tilde \epsilon}^h_i
\sim | ( {\tilde \Delta}_{qQ}^\prime )_{ai} |/m_Q \sim m_{q_i} / m_Q .
\end{equation}
By considering these arguments, the case of the basis (a) may be regarded
as a special case of the basis (b).
It is, in particular, interesting that the specific relation
(\ref{eqn:hq-lmq}) for the Yukawa couplings,
which may be invoked ad hoc in the basis (b),
is obtained naturally starting from the basis (a).

We next consider the cases with $ N_Q = 3 $ including the seesaw model.
The quark mass matrix in the basis (a)
can be deformed to the seesaw form
by the exchange $ q^c \leftrightarrow Q^c $:
\begin{equation}
\left( \begin{array}{cc}
M_q & \Delta_{qQ} \\ {\bf 0} & M_Q \end{array} \right)
\rightarrow
\left( \begin{array}{cc} {\bf 0} & M_Q \\
M_q & \Delta_{qQ} \end{array} \right) .
\end{equation}
Then, if $ | \Delta_{qQ} | \gtrsim | M_Q | $,
the relations for the $ q $-$ Q $ mixing effects in the seesaw case
can be applied.
We have also seen that if $ | \Delta_{qQ} | \gtrsim | M_Q | $
in the basis (a), it is appropriate to transfer
from the basis (a) to the basis (b).
Hence, it is expected that the seesaw case is even related
to the case of basis (b).
In fact, the seesaw quark mass matrix can be deformed as
\begin{equation}
\left( \begin{array}{cc}
{\bf 0} & \Delta_{qQ} \\ \Delta_{qQ}^\prime & M_Q \end{array} \right)
\rightarrow
\left( \begin{array}{cc} {\tilde M}_q & {\bf 0} \\
{\tilde \Delta}_{qQ}^\prime & {\tilde M}_Q \end{array} \right) .
\end{equation}
Here, the $ \Delta_{qQ} $ term is first eliminated
by the right-handed quark transformation,
and then the $ {\tilde M}_q $ term is made diagonal
by the ordinary quark transformation.
The relation between $ {\tilde \Delta}_{qQ}^\prime $ and $ {\tilde M}_q $,
which is similar to Eq. (\ref{eqn:DqQp-Mq}), is obtained
since they both stem from the $ \Delta_{qQ}^\prime $ term.

If we start with a quark basis with general form of $ {\cal M}_{\cal Q} $,
the quark masses and mixings are apparently given as a mixture
of the three representative cases.
The quark masses are obtained as
\[
m_{q_i} \sim \lambda_{q_i} v + \epsilon^f_i \epsilon^h_i m_Q .
\]
Then, the hierarchy of the ordinary quark masses may be reproduced
by the tuning between these two contributions
or by relating the $ q $-$ Q $ mixing parameters
$ \epsilon^f_i $ and/or $ \epsilon^h_i $
to the corresponding Yukawa couplings $ \lambda_{q_i} $.
The latter choice seems to be technically natural.
The $ q $-$ Q $ mixing effects on the ordinary quark mixing
are mainly described in terms of the ratios $ \epsilon^f_i / \epsilon^f_j $
and $ \epsilon^h_i / \epsilon^h_j $,
as given in Eqs. (\ref{eqn:VqL-sss}) and (\ref{eqn:VqR-sss})
for the seesaw case.  It should anyway be remarked
that these $ q $-$ Q $ mixing effects are reduced
to those given in one of the three representative bases,
if appropriate rearrangements are made for the quark fields
and relevant Yukawa couplings.

To summarize, we may take complementarily
the basis (a) with $ \Delta_{qQ}^\prime = {\bf 0} $,
the basis (b) with $ \Delta_{qQ} = {\bf 0} $
and the seesaw with $ M_q = {\bf 0} $.
Some choice of the basis may appear to be more suitable than the others,
depending on the parameter ranges
and also the symmetries and matter contents.
For example, the seesaw form mass matrix
is obtained readily in the left-right gauge models
\cite{qseesaw,BM,ssinverted}.
The basis (b) will be suitable if $ | \Delta_{qQ} | \gg | M_Q | $
in the basis (a).
On the other hand, the ad hoc relation (\ref{eqn:hq-lmq}) in the basis (b)
is realized naturally starting from the basis (a).
In the model with complex singlet Higgs field $ S $,
the condition $ \Delta_{qQ} = {\bf 0} $ in the basis (b) might require
a tuning between the $ f_Q $ and $ f_Q^\prime $ couplings,
as seen in Eq. (\ref{eqn:fQ0}), unless $ f_Q^\prime \equiv {\bf 0} $
is ensured by means of certain symmetry.
In any case, by taking the appropriate quark basis
we can find the reasonable regions of the model parameter space
where the actual masses of the ordinary quarks
are reproduced even in the presence of singlet quarks.

\section{Flavor changing interactions}
\label{sec:FCI}

In this section, we examine the flavor changing interactions of quarks
which are affected by the $ q $-$ Q $ mixing.
Specifically, the CKM unitarity within the ordinary quark sector
is violated, and the FCNC's arise both in the gauge and scalar couplings
\cite{AgBr,LL,BB,qQmix,KY,Fr}.
We would like to clarify the structures
of these flavor changing interactions
for the representative bases.  In fact, they are described appropriately
in terms of the $ q $-$ Q $ mixing parameters and the quark masses.
Hence, the present model with singlet quarks
may provide an interesting extension
of the notion of natural flavor conservation
\cite{GWP}.

\subsection{Charged currents}

The charged gauge interaction coupled to the $ W $ boson
is expressed in terms of the quark mass eigenstates as
\begin{eqnarray}
{\cal L}_{\rm CC}(W)
&=& g W^+_\mu {\cal U}^\dagger \sigma^\mu {\cal V} {\cal D} + {\rm h.c.}
\nonumber \\
&=& g W^+_\mu u^\dagger \sigma^\mu V d + \cdots .
\end{eqnarray}
Here the generalized left-handed quark mixing matrix for the charged
weak currents is given by
\begin{equation}
{\cal V} = {\cal V}_{{\cal U}_{\rm L}}^\dagger
\left( \begin{array}{cc} V_0 & {\bf 0} \\ {\bf 0} & {\bf 0}
\end{array} \right) {\cal V}_{{\cal D}_{\rm L}} .
\end{equation}
The CKM matrix $ V $ for the ordinary quarks,
which is included in $ {\cal V} $ as a submatrix,
is actually modified from the original $ V_0 $ as
\begin{equation}
V = V_{u_{\rm L}}^\dagger V_0 V_{d_{\rm L}} .
\label{eqn:VCKM}
\end{equation}
Here, $ V_{u_{\rm L}} $ and $ V_{d_{\rm L}} $ are the left-handed
ordinary quark mixings induced by the $ q $-$ Q $ mixing,
which are presented in the previous section.

In the present model containing only one Higgs doublet,
there is no physical charged Higgs particle mediating
the scalar interactions of quarks.
If supersymmetric models with a pair of Higgs doublets
$ H_1 $ and $ H_2 $ are considered,
one physical charged scalar particle appears.
Then, this charged scalar particle as well as
the Nambu-Goldstone mode absorbed by the $ W $ boson
have the Yukawa couplings which are described
in terms of the quark masses and CKM matrix as
\begin{equation}
\Lambda_u^+ = ( {\bar M}_u / \langle H_2^0 \rangle ) V , \
\Lambda_d^- = ( {\bar M}_d / \langle H_1^0 \rangle ) V^\dagger .
\end{equation}
Therefore, these charged scalar couplings 
with the same CKM structure as the charged gauge interaction
do not provide so distinct effects on the flavor changing processes.

The unitarity violation of the CKM matrix $ V $ is calculated
with the unitarity relations (\ref{eqn:unvVq1}) and (\ref{eqn:unvVq2})
of $ {\cal V}_{{\cal Q}_{\rm L}} $:
\begin{eqnarray}
V^\dagger V - {\bf 1}
&=& - \epsilon_{d_{\rm L}}^\prime \epsilon_{d_{\rm L}}^{\prime \dagger}
- V_{d_{\rm L}}^\dagger V_0^\dagger
\epsilon_{u_{\rm L}} \epsilon_{u_{\rm L}}^\dagger V_0 V_{d_{\rm L}} ,
\label{eqn:unvVCKM1} \\
V V^\dagger - {\bf 1}
&=& - \epsilon_{u_{\rm L}}^\prime \epsilon_{u_{\rm L}}^{\prime \dagger}
- V_{u_{\rm L}}^\dagger V_0
\epsilon_{d_{\rm L}} \epsilon_{d_{\rm L}}^\dagger V_0^\dagger V_{u_{\rm L}} .
\label{eqn:unvVCKM2}
\end{eqnarray}
As shown in the next subsection,
the modification of the $ Z $ mediated neutral currents
as well as this CKM unitarity violation
are described in terms of the second order $ q $-$ Q $ mixing factors
$ \epsilon_{q_{\rm L}} \epsilon_{q_{\rm L}}^\dagger $ and
$ \epsilon_{q_{\rm L}}^\prime \epsilon_{q_{\rm L}}^{\prime \dagger} $
\cite{AgBr,LL,BB,qQmix,KY,Fr}.
In the basis (a) with $ \Delta_{qQ}^\prime = {\bf 0} $,
the CKM unitarity violation is actually far below the experimental bounds
\cite{ParticleData}.
This is due to the fact that the left-handed $ q $-$ Q $ mixing
is suppressed by the $ q / Q $ mass ratios, as seen
in Eq. (\ref{eqn:eqL-a}).
On the other hand, if the model parameters are taken
so that the basis (b) with $ \Delta_{qQ} = {\bf 0} $ is relevant,
the $ q $-$ Q $ mixing effects on the flavor changing interactions
are not necessarily suppressed by the $ q / Q $ mass ratios.
Hence, they can be comparable to the current experimental bounds,
as usually considered in the literature
\cite{LL,qQmix}.
As for the seesaw model, the $ q $-$ Q $ mixing parameters
may be related to the ordinary quark masses,
as discussed in the previous section.
Then, the $ q $-$ Q $ mixing effects are substantially suppressed.

We now consider how the CKM structure
with small flavor changing elements is reproduced
in the presence of $ q $-$ Q $ mixing.
We have seen that in some cases the left-handed ordinary quark mixing
$ V_{q_{\rm L}} $ induced by the $ q $-$ Q $ mixing is related
to the ordinary quark mass hierarchy.
Then, $ V_{q_{\rm L}} $ is close enough to the unit matrix,
and the actual CKM matrix $ V $ can be obtained readily
by taking $ V_0 \simeq V $.
In other cases, as seen for example in Eq. (\ref{eqn:qQmix-ii}),
$ V_{q_{\rm L}} $ itself may deviate significantly from the unit matrix.
Even in such cases, the unitarity violation of $ V_{q_{\rm L}} $
arising at the second order of the $ q $-$ Q $ mixing parameters
are constrained to be small enough phenomenologically.
Then, the realistic CKM matrix can be reproduced
by taking $ V_0 \simeq V_{u_{\rm L}} V V_{d_{\rm L}}^\dagger $.
As long as the mixing matrix $ V_0 $ can be taken freely,
it is in fact impossible to make some definite predictions
on the CKM matrix.
It is, however, at least technically natural
that the actual CKM matrix is close enough to the unit matrix.
This choice for the CKM matrix can be ensured
by means of the approximate flavor symmetries,
as is the case in the minimal standard model.
It will be an interesting possibility,
as considered in the seesaw models for quark masses
\cite{qseesaw},
that some predictions on the CKM matrix are obtained
by invoking some global chiral symmetry to restrict the forms
of Yukawa coupling matrices.

\subsection{Neutral currents}

We next describe the neutral currents of quarks
coupled to the $ Z $ boson and Higgs scalar particles,
which are also modified by the $ q $-$ Q $ mixing.

\subsubsection{Neutral gauge couplings}

The neutral gauge interaction of quarks mediated by the $ Z $ boson
is given by
\begin{equation}
{\cal L}_{\rm NC}(Z)
= \frac{g}{\cos \theta_W} Z_\mu J_Z^\mu
\end{equation}
with
\begin{equation}
J_Z^\mu = \sum_{{\cal U},{\cal D}}
{\cal Q}^\dagger \sigma^\mu {\cal Z}_{\cal Q} {\cal Q} \
+ \sum_{{\cal U}^c ,{\cal D}^c}
{\cal Q}^{c \dagger} \sigma^\mu {\cal Z}_{{\cal Q}^c} {\cal Q}^c .
\end{equation}
The coupling matrices are given by
\begin{eqnarray}
{\cal Z}_{\cal Q} &=& {\cal V}_{{\cal Q}_{\rm L}}^\dagger
{\cal Z}_{\cal Q}^0 {\cal V}_{{\cal Q}_{\rm L}} ,
\\
{\cal Z}_{{\cal Q}^c} &=& I_Z (q^c_0)
\left( \begin{array}{cc} {\bf 1} & {\bf 0} \\
{\bf 0} & {\bf 1} \end{array} \right) ,
\end{eqnarray}
where
\begin{equation}
{\cal Z}_{\cal Q}^0 = \left( \begin{array}{cc}
I_Z (q_0) {\bf 1} & {\bf 0} \\
{\bf 0} & I_Z (Q_0) {\bf 1} \end{array} \right) ,
\end{equation}
and
\begin{equation}
I_Z ({\cal F}) = I_3 ({\cal F})
- \sin^2 \theta_W Q_{\rm em} ({\cal F})
\end{equation}
for $ {\cal F} = q_0 , q^c_0 , Q_0 , Q^c_0 $.

The right-handed couplings are unchanged
with the diagonal and flavor-universal form
for $ I_3 ( q^c_0 ) = I_3 ( Q^c_0 ) = 0 $.
The variation of the left-handed couplings induced
by the $ q $-$ Q $ mixing is calculated as
\begin{eqnarray}
\Delta {\cal Z}_{\cal Q}
& \equiv & {\cal Z}_{\cal Q} - {\cal Z}_{\cal Q}^0
\nonumber \\
&=& I_3 ( q_0 ) \left( \begin{array}{cc}
- \epsilon_{q_{\rm L}}^\prime \epsilon_{q_{\rm L}}^{\prime \dagger}
     & V_{q_{\rm L}}^\dagger \epsilon_{q_{\rm L}} \\
\epsilon_{q_{\rm L}}^\dagger V_{q_{\rm L}}
     & \epsilon_{q_{\rm L}}^\dagger \epsilon_{q_{\rm L}}
\end{array} \right) .
\end{eqnarray}
The upper-left component, say $ \Delta {\cal Z}_{\cal Q} [q] $,
in particular, describes the neutral currents among the ordinary quarks:
\begin{equation}
\Delta {\cal Z}_{\cal Q} [q]
= - I_3 (q_0) \epsilon_{q_{\rm L}}^\prime
\epsilon_{q_{\rm L}}^{\prime \dagger} .
\label{eqn:DZQ}
\end{equation}
It should here be noticed that this variation of the neutral currents
(\ref{eqn:DZQ}) as well as the CKM unitarity violation
given in Eqs. (\ref{eqn:unvVCKM1}) and (\ref{eqn:unvVCKM2})
arise at the second order of $ q $-$ Q $ mixing
\cite{AgBr,LL,BB,qQmix,KY,Fr}.

In the basis (a) with $ \Delta_{qQ}^\prime = {\bf 0} $,
the variation of the $ Z $ mediated neutral currents
induced by the $ q $-$ Q $ mixing is estimated with Eq. (\ref{eqn:eqL-a}) as
\begin{equation}
\Delta {\cal Z}_{\cal Q} [q]_{ij}
\sim ( m_{q_i} / m_Q ) ( m_{q_j} / m_Q ) \epsilon^f_i \epsilon^f_j .
\label{eqn:DZQ-a}
\end{equation}
This correction as well as the CKM unitarity violation
are suppressed substantially by the second order of $ q/Q $ mass ratios,
providing negligible effects except for those involving the top quark.

On the other hand, in the basis (b) with $ \Delta_{qQ} = {\bf 0} $
we obtain
\begin{equation}
\Delta {\cal Z}_{\cal Q} [q]_{ij}
\sim \epsilon^h_i \epsilon^h_j .
\label{eqn:DZQ-b}
\end{equation}
This $ q $-$ Q $ mixing effect on the $ Z $ mediated neutral currents
as well as the CKM unitarity violation
are no longer suppressed by the $ q/Q $ mass ratios.
Then, some meaningful constraints are placed phenomenologically
on the $ q $-$ Q $ mixing, and such constraints provide restrictions
on the possible contributions to the flavor changing processes
\cite{LL,qQmix}.

The modifications of the neutral gauge couplings
as given in Eq. (\ref{eqn:DZQ-b}) for the basis (b)
are also obtained for the seesaw model
with $ m_{Q_1} , m_{Q_2} , m_{Q_3} \sim m_Q $.
Then, for the typical cases considered
in Eqs. (\ref{eqn:qQmix-i}), (\ref{eqn:qQmix-ii}) and (\ref{eqn:qQmix-iii}),
they are estimated as
\begin{eqnarray}
{\mbox{(i)}} &:& \Delta {\cal Z}_{\cal Q} [q]_{ij}
\sim {\sqrt{( m_{q_i} / m_Q ) ( m_{q_j} / m_Q )}} ,
\label{eqn:DZQ-i} \\
{\mbox{(ii)}} &:&
\Delta {\cal Z}_{\cal Q} [q]_{ij}
\sim ( {\bar h} v / m_Q )^2 ,
\label{eqn:DZQ-ii} \\
{\mbox{(iii)}} &:& \Delta {\cal Z}_{\cal Q} [q]_{ij}
\sim  ( m_{q_i} / {\bar f} v_S ) ( m_{q_j} / {\bar f} v_S ) .
\label{eqn:DZQ-iii}
\end{eqnarray}
Here, for the case (i) we have estimates
with the ordinary quark masses as
\begin{eqnarray}
\Delta {\cal Z}_{\cal U} [u]
& \sim & \left( \begin{array}{ccc}
10^{-5} & 10^{-4} & 10^{-3} \\
10^{-4} & 10^{-3} & 10^{-2} \\
10^{-3} & 10^{-2} & 10^{-1} \end{array} \right)  \frac{500 {\rm GeV}}{m_U} ,
\\
\Delta {\cal Z}_{\cal D} [d]
& \sim & \left( \begin{array}{ccc}
10^{-5} & 10^{-5} & 10^{-4} \\
10^{-5} & 10^{-4} & 10^{-2} \\
10^{-4} & 10^{-2} & 10^{-2} \end{array} \right) \frac{500 {\rm GeV}}{m_D} .
\end{eqnarray}
These $ q $-$ Q $ mixing effects in the case (i) 
can be comparable to the experimental bounds
\cite{ParticleData}
for $ m_Q \sim 500 {\rm GeV} $.
In the case (ii), where the $ Z $ mediated FCNC's are
rather flavor-independent, some stringent constraints
will be placed phenomenologically on the mass ratio $ {\bar h} v / m_Q $.

As for the seesaw model with $ m_{Q_1} \gg m_{Q_2} \gg m_{Q_3} $,
where the left-handed $ q $-$ Q $ mixing is given
by Eq. (\ref{eqn:eqL-ssi}), the $ Z $ mediated FCNC's appear to be
the same as given in Eq. (\ref{eqn:DZQ-iii}) for the case (iii).
They are fairly suppressed by the second order of the ordinary quark masses,
which is also the case in the basis (a).

\subsubsection{Neutral scalar couplings}

The neutral scalar couplings of quarks are extracted
from Eq. (\ref{eqn:LYukawa}) as
\begin{equation}
{\cal L}_{\rm NC}( \phi )
= - \frac{1}{\sqrt 2} \sum_{\alpha = 0,1,2}
{\cal Q}^c \Lambda_{\cal Q}^\alpha {\cal Q} \phi_\alpha \ + \ {\rm h.c.},
\end{equation}
where $ \phi_0 $, $ \phi_1 $, $ \phi_2 $ represent the mass eigenstates
of the neutral scalar fields.
The coupling matrices are given by
\begin{equation}
\Lambda_{\cal Q}^\alpha
= O_{\alpha 0} \Lambda_{\cal Q}^H
+ O_{\alpha 1} \Lambda_{\cal Q}^{S_+}
+ i O_{\alpha 2} \Lambda_{\cal Q}^{S_-}
\end{equation}
with
\begin{eqnarray}
\Lambda_{\cal Q}^H &=& {\cal V}_{{\cal Q}_{\rm R}}^\dagger
\left( \begin{array}{cc} \lambda_q & {\bf 0} \\
h_q & {\bf 0} \end{array} \right)
{\cal V}_{{\cal Q}_{\rm L}} ,
\\
\Lambda_{\cal Q}^{S_\pm} &=& {\cal V}_{{\cal Q}_{\rm R}}^\dagger
\left( \begin{array}{cc}
{\bf 0} & f_Q \pm f^\prime_Q \\
{\bf 0} & \lambda_Q \pm \lambda^\prime_Q
\end{array} \right) {\cal V}_{{\cal Q}_{\rm L}} .
\end{eqnarray}
Here an orthogonal matrix $ O $ is introduced
to parametrize the mass eigenstates of the neutral scalar fields:
\begin{equation}
\left( \begin{array}{c} \phi_0 \\ \phi_1 \\ \phi_2 \end{array} \right)
= O \left( \begin{array}{c} h_1 \\ s_1 \\
s_2 \end{array} \right) .
\end{equation}
The original complex Higgs fields are decomposed
with the real scalar fields as
\begin{eqnarray}
H^0 &=& \langle H^0 \rangle + ( h_1 + i h_2 )/{\sqrt 2} ,
\\
S &=& \langle S \rangle + ( s_1 + i s_2 )/{\sqrt 2} .
\end{eqnarray}
While the Nambu-Goldstone mode $ h_2 $ is absorbed by the $ Z $ boson,
the remaining $ h_1 $, $ s_1 $, $ s_2 $ are combined
to form the mass eigenstates $ \phi_\alpha $.
At present, the masses $ m_{\phi_\alpha} $
and mixing matrix $ O $ of the neutral scalar fields
should be regarded as free parameters varying in some reasonable range.
If the hierarchy $ v_S \gg v $ is realized,
the mixing between the Higgs doublet and singlet
will be of the order of $ v/v_S $.

Let us examine in detail the structures
of the submatrices, say $ \Lambda_{\cal Q}^\alpha [q] $,
describing the neutral scalar couplings of the ordinary quarks:
\begin{equation}
\Lambda_{\cal Q}^\alpha [q] = O_{\alpha 0} \Lambda_{\cal Q}^H [q]
+ O_{\alpha 1} \Lambda_{\cal Q}^{S_+} [q]
+ i O_{\alpha 2} \Lambda_{\cal Q}^{S_-} [q] ,
\label{eqn:LmQ}
\end{equation}
where
\begin{equation}
\Lambda_{\cal Q}^H [q] = {\hat \lambda}_q + {\hat h}_q , \
\Lambda_{\cal Q}^{S_\pm} [q] = {\hat f}_Q^\pm + {\hat \lambda}_Q^\pm
\label{eqn:LmQ-HS}
\end{equation}
with
\begin{eqnarray}
{\hat \lambda}_q &=&
V_{q_{\rm R}}^\dagger \lambda_q V_{q_{\rm L}} ,
\label{eqn:lmqtld} \\
{\hat h}_q &=&
- \epsilon_{q_{\rm R}}^\prime h_q V_{q_{\rm L}} ,
\label{eqn:hqtld} \\
{\hat f}_Q^\pm &=&
- V_{q_{\rm R}}^\dagger ( f_Q \pm f_Q^\prime )
\epsilon_{q_{\rm L}}^{\prime \dagger} ,
\label{eqn:fQtld} \\
{\hat \lambda}_Q^\pm &=&
\epsilon_{q_{\rm R}}^\prime ( \lambda_Q \pm \lambda_Q^\prime )
\epsilon_{q_{\rm L}}^{\prime \dagger} .
\label{eqn:lmQtld}
\end{eqnarray}
By considering the relations for the $ q $-$ Q $ mixing effects,
which are described in Sec. \ref{sec:mass-mixing}, the flavor structure
of the neutral scalar couplings $ \Lambda_{\cal Q}^\alpha [q]_{ij} $
is specified for the respective bases.
We suppress below for simplicity the neutral Higgs mixing parameters
by assuming $ O_{\alpha \beta} \sim 1 $,
though they are readily recovered for the scalar couplings.

We first note the relations,
\begin{eqnarray}
{\hat \lambda}_q + {\hat h}_q &=& \Lambda_{\cal Q}^H [q]
= ( {\bar M_q} / v ) V_{q_{\rm L}}^\dagger V_{q_{\rm L}} ,
\label{eqn:lmqhqtld} \\
{\hat \lambda}_q + ( v_S / v ) {\hat f}_Q^S
&=& V_{q_{\rm R}}^\dagger V_{q_{\rm R}} ( {\bar M_q} / v ) ,
\label{eqn:lmqfQStld} \\
( v_S / v ) {\hat \lambda}_Q^S + {\hat h}_q
&=& \epsilon_{q_{\rm R}}^\prime \epsilon_{q_{\rm R}}^{\prime \dagger}
( {\bar M_q} / v ) ,
\label{eqn:lmQShqtld} \\
( v_S / v ) ( {\hat \lambda}_Q^S + {\hat f}_Q^S )
&=& ( {\bar M_q} / v )
\epsilon_{q_{\rm L}}^\prime \epsilon_{q_{\rm L}}^{\prime \dagger} ,
\label{eqn:lmQSfQStld}
\end{eqnarray}
where
\begin{eqnarray}
{\hat f}_Q^S &=& - V_{q_{\rm R}}^\dagger
( f_Q {\rm e}^{i \phi_S} + f^\prime_Q {\rm e}^{-i \phi_S} )
\epsilon_{q_{\rm L}}^{\prime \dagger} ,
\\
{\hat \lambda}_Q^S &=&
\epsilon_{q_{\rm R}}^\prime
( \lambda_Q {\rm e}^{i \phi_S} + \lambda^\prime_Q {\rm e}^{-i \phi_S} )
\epsilon_{q_{\rm L}}^{\prime \dagger} .
\end{eqnarray}
In order to derive these relations, Eq. (\ref{eqn:MQdiagonal})
is multiplied by the products of matrices,
\[
{\cal V}_{{\cal Q}_\chi}^\dagger
\left( \begin{array}{cc} {\bf 1} & {\bf 0} \\ {\bf 0} & {\bf 0}
\end{array} \right) {\cal V}_{{\cal Q}_\chi} , \
{\cal V}_{{\cal Q}_\chi}^\dagger
\left( \begin{array}{cc} {\bf 0} & {\bf 0} \\ {\bf 0} & {\bf 1}
\end{array} \right) {\cal V}_{{\cal Q}_\chi} ,
\]
from the right for $ \chi = {\rm L} $ or left for $ \chi = {\rm R} $.
The relation (\ref{eqn:lmqhqtld}), in particular, implies physically
that the couplings of the $ Z $ boson and the Higgs field
$ H^0 $ including the Nambu-Goldstone mode
have the same flavor structure.
In fact, by considering Eqs. (\ref{eqn:DZQ}) and (\ref{eqn:lmqhqtld})
with Eq. (\ref{eqn:unvVq1}) we find the relation for the FCNC's,
\begin{equation}
( \Lambda_{\cal Q}^H [q] )_{ij}^{( i \not= j )}
= ( m_{q_i} / v ) \Delta {\cal Z}_{\cal Q} [q]_{ij} / I_3 (q_0) .
\label{eqn:lmqhqtld-DZQ}
\end{equation}
Then, it is sufficient to calculate the contributions
$ \Lambda_{\cal Q}^{S_\pm} [q] $ with the singlet Higgs $ S $.

In the basis (a) with $ \Delta_{qQ}^\prime = {\bf 0} $,
the $ Z $ mediated FCNC's are suppressed
by the second order of $ q/Q $ mass ratios.
Hence, the $ f_Q $ and $ f_Q^\prime $ couplings
provide dominant contributions to the scalar FCNC's.
They are estimated as
\begin{eqnarray}
( {\hat f}_Q^\pm )_{ij}
&=& - ( V_{q_{\rm R}}^\dagger )_{ii}
( f_Q^\pm )_{ia} ( \epsilon_{q_{\rm L}}^{\prime \dagger} )_{aj}
\nonumber \\
&-&
\sum_{k \not= i} ( V_{q_{\rm R}}^\dagger )_{ik}
( f_Q^\pm )_{ka} ( \epsilon_{q_{\rm L}}^{\prime \dagger} )_{aj} ,
\nonumber
\end{eqnarray}
where Eqs. (\ref{eqn:eqL-a}) and (\ref{eqn:VqR-a})
for the quark mixings are considered.
The first term amounts to the order of
$ ( m_{q_j} / v_S ) \epsilon^f_i \epsilon^f_j $.
The second term is estimated as
$ \epsilon^f_i \epsilon^f_k \epsilon^f_k ( m_Q / v_S)
( m_{q_j} / m_Q ) \epsilon^f_j
\lesssim ( m_{q_j} / v_S ) \epsilon^f_i \epsilon^f_j $.
Similar estimates are made for the $ \lambda_Q $ and $ \lambda_Q^\prime $
contributions.  Then, the leading contributions
to the neutral scalar couplings in the basis (a)
with $ \Delta_{qQ}^\prime = {\bf 0} $ are given by
\begin{equation}
\Lambda_{\cal Q}^\alpha [q]_{ij} \sim ( m_{q_i} / v ) \delta_{ij}
+ ( m_{q_j} / v_S ) \epsilon^f_i \epsilon^f_j .
\label{eqn:LmQ-a}
\end{equation}
Here, in contrast to the $ Z $ mediated FCNC's,
the scalar FCNC's in the basis (a) are suppressed
only by the first order of ordinary quark masses.
The relevant factors are estimated as
\begin{eqnarray}
\frac{m_{u_j}}{v_S}  & \sim &
\left( 10^{-5} , 10^{-3} , 10^{-1} \right) \frac{500{\rm GeV}}{v_S} ,
\label{eqn:mu-vS} \\
\frac{m_{d_j}}{v_S} & \sim &
\left( 10^{-5} , 10^{-4} , 10^{-2} \right) \frac{500{\rm GeV}}{v_S} .
\label{eqn:md-vS}
\end{eqnarray}
Then, these scalar FCNC's are expected to provide
significant phenomenological effects
for $ \epsilon^f_i \sim 0.1 - 1 $
and $ v_S \sim 100 {\rm GeV} - 1 {\rm TeV} $
\cite{KY}.

We here mention that in certain models these contributions
to the scalar FCNC's substantially cancel out.
For instance, suppose that $ f_Q^\prime = {\bf 0} $
and $ \lambda_Q^\prime = {\bf 0} $, as is the case
for the one real $ S $ model and the one supersymmetric $ S $ model.
Then, we have $ {\hat f}_Q + {\hat \lambda}_Q
= {\rm e}^{-i \phi_S} ( {\hat f}_Q^S + {\hat \lambda}_Q^S )
= ( {\bar M_q} / v_S )
\epsilon_{q_{\rm L}}^\prime \epsilon_{q_{\rm L}}^{\prime \dagger} $
from Eq. (\ref{eqn:lmQSfQStld}), which is even smaller
by the factor $ v / v_S $ than the $ \Lambda_{\cal Q}^H $ contribution
given in Eq. (\ref{eqn:lmqhqtld}).
Hence, for this specific case
with $ f_Q^\prime = {\bf 0} $ and $ \lambda_Q^\prime = {\bf 0} $
we have the scalar FCNC's
$ \Lambda_{\cal Q}^\alpha [q]_{ij}^{( i \not= j )}
\sim ( m_{q_i} / v ) ( m_{q_i} / m_Q ) ( m_{q_j} / m_Q )
\epsilon^f_i \epsilon^f_j $,
which are related to the $ Z $ mediated FCNC's
as seen in Eq. (\ref{eqn:lmqhqtld-DZQ}). 
This result is also valid for the no singlet Higgs $ S $ model with
bare $ \Delta_{qQ} $ and $ M_Q $ terms.
It should further be remarked for completeness
that in some models only $ M_Q $ is the bare mass term,
but the $ \Delta_{qQ} $ term is provided by the singlet Higgs $ S $
either real or complex.
In this case, the above cancellation between
the $ f_Q $ and $ \lambda_Q $ couplings does not take place,
and hence the scalar FCNC's are still given by Eq. (\ref{eqn:LmQ-a}).

We next consider the basis (b) with $ \Delta_{qQ} = {\bf 0} $.
It should be noted that the $ f_Q $ and $ f_Q^\prime $ couplings
may in general take some nonzero values as
\begin{equation}
| ( f_Q )_{ia} | , | ( f_Q^\prime )_{ia} | \sim {\bar f}_i .
\end{equation}
Then, even though the specific combination $ {\hat f}_Q^S $
vanishes due to the condition $ \Delta_{qQ} = {\bf 0} $,
there is no reason to have cancellation between
the $ f_Q $ and $ f_Q^\prime $ contributions in Eq. (\ref{eqn:fQtld}).
Hence, we obtain
\begin{equation}
\Lambda_{\cal Q}^\alpha [q]_{ij}
\sim ( m_Q / v_S ) {\bar f}_i \epsilon^h_j
+ ( m_{q_i} / v ) ( \delta_{ij} + \epsilon^h_i \epsilon^h_j ) .
\label{eqn:LmQ-b}
\end{equation}
The first term from the $ f_Q $ and $ f_Q^\prime $ couplings
is no longer suppressed by the ordinary quark masses.
The flavor changing part of the second term from $ \Lambda_{\cal Q}^H [q] $
is related to the $ Z $ mediated FCNC's,
as given in Eq. (\ref{eqn:lmqhqtld-DZQ}).
The $ \lambda_Q $ and $ \lambda_Q^\prime $ contributions
in Eq. (\ref{eqn:lmQtld}) are also estimated as
$ ( m_{q_i} / v_S ) \epsilon^h_i \epsilon^h_j $
with Eqs. (\ref{eqn:eqL-b}) and (\ref{eqn:eqR-b}).
It should here be remembered that in some models
the $ f_Q^\prime $ coupling is absent.
Then, the $ f_Q $ coupling is eliminated for $ \Delta_{qQ} = {\bf 0} $,
and hence the first term disappears in Eq. (\ref{eqn:LmQ-b}).

The quark mass matrix of the seesaw form may be deformed formally
to those in the bases (a) and (b),
as seen in Sec. \ref{sec:mass-mixing}.
Hence, similar features are expected for the neutral scalar couplings,
which have been observed so far.
For definiteness, we consider the case
where the $ f_Q^\prime $ coupling is absent
and the submatrix $ M_Q $ is a bare mass term.
Then, the $ h_q $ and $ f_Q $ contributions are determined,
respectively, from Eqs. (\ref{eqn:lmqhqtld}) and (\ref{eqn:lmqfQStld})
with $ \lambda_q = {\bf 0} $ and $ f_Q^\prime = {\bf 0} $ as
\begin{equation}
\Lambda_{\cal Q}^\alpha [q]_{ij}
\sim ( V_{q_{\rm R}}^\dagger V_{q_{\rm R}} )_{ij} ( m_{q_j} / v_S )
+ ( m_{q_i} / v ) ( V_{q_{\rm L}}^\dagger V_{q_{\rm L}} )_{ij} .
\label{eqn:LmQ-ss}
\end{equation}
Here, the scalar FCNC's ($ i \not= j $) are described
in terms of the ordinary quark masses
and the unitarity violation (\ref{eqn:unvVq1})
of the ordinary quark mixings $ V_{q_{\rm L}} $ and $ V_{q_{\rm R}} $
induced by the $ q $-$ Q $ mixing.
In particular, the second term is related to the $ Z $ mediated FCNC's.
In the case of $ m_{Q_1} , m_{Q_2} , m_{Q_3} \sim m_Q $,
the first term can be significant with the form as given
in Eq. (\ref{eqn:LmQ-a})
for the basis (a).  On the other hand, if the inverted hierarchy
$ m_{Q_1} \gg m_{Q_2} \gg m_{Q_3} $ is realized,
these scalar FCNC's are related to the ordinary quark masses
with Eqs. (\ref{eqn:eqL-ssi}) and (\ref{eqn:eqR-ssi}).
Hence, they appear to be negligibly small.

\subsubsection{FCNC's ($ Z $) versus FCNC's ($ \phi $)}

We have examined so far the structures of the FCNC's
for the representative bases, which are summarized
in Table \ref{table:FCNC's}.
We here, in particular, note that in some cases
the scalar FCNC's ($ \phi $) can be much larger than
the $ Z $ mediated FCNC's ($ Z $).

In the basis (a) with $ \Delta_{qQ}^\prime = {\bf 0} $,
the FCNC's ($ \phi $) arise at the first order of
$ q/Q $ mass ratios, while the FCNC's ($ Z $) are fairly suppressed
by the second order of $ q/Q $ mass ratios.
Hence, as seen in Eqs. (\ref{eqn:LmQ-a}),
(\ref{eqn:mu-vS}) and (\ref{eqn:md-vS}),
the FCNC's ($ \phi $) are expected to provide
significant physical effects
for $ m_Q , m_{\phi_\alpha} \sim 100 {\rm GeV} - 1 {\rm TeV} $.

In the basis (b) with $ \Delta_{qQ} = {\bf 0} $,
as seen in Eq. (\ref{eqn:LmQ-b}),
if both the $ f_Q $ and $ f_Q^\prime $ couplings
are present with complex $ S $ or several real $ S $'s,
the FCNC's ($ \phi $) contains the term
which is not related to the FCNC's ($ Z $).
Then, the contributions of FCNC's ($ \phi $) may exceed
those of FCNC's ($ Z $) if the $ f_Q $ and $ f_Q^\prime $ couplings
are large enough with
$ m_Q , m_{\phi_\alpha} \sim 100 {\rm GeV} - 1 {\rm TeV} $.

The seesaw model with $ m_{Q_1} , m_{Q_2} , m_{Q_3} \sim m_Q $
has a hybrid feature of the above two cases for the FCNC's, i.e.,
the FCNC's ($ \phi $) has the structure the same as in the basis (a),
while the FCNC's ($ Z $) the same as in the basis (b).
In the seesaw model with $ m_{Q_1} \gg m_{Q_2} \gg m_{Q_3} $,
the FCNC's ($ Z $) and FCNC's ($ \phi $) are both negligibly small
suppressed by the powers of the ordinary quark masses.

We would anyway like to emphasize
that in some cases the effects of the FCNC's ($ \phi $)
can be more important than those of the FCNC's ($ Z $).
Then, the neutral Higgs contributions
to the flavor changing and $ CP $ violating processes
may rather serve as signals for the new physics beyond the standard model
\cite{BB,KY,BM}.
This possibility has not been paid so much attention before
in the models with singlet quarks.

\section{Numerical analysis}
\label{sec:numerical}

We here perform a detailed numerical analysis
for calculating the quark mixings and flavor changing couplings
which are induced by the $ q $-$ Q $ mixing.
The flavor structures of these $ q $-$ Q $ mixing effects
have been described in the previous sections.
They are really confirmed by this numerical analysis.

We begin with taking some reasonable values for the model parameters.
The VEV's of the Higgs fields are taken typically as
\[
v = 246 {\rm GeV} \ , \ v_S = 500 {\rm GeV} \ .
\]
The singlet quark masses are chosen as
\[
m_{Q_a} \sim 300 {\rm GeV} - 1 {\rm TeV} \ ( a = 1, 2, \ldots , N_Q )
\]
(except for the seesaw model with $ m_{Q_1} \gg m_{Q_2} \gg m_{Q_3} $).
This is made by taking suitably the $ \lambda_Q $ and $ \lambda_Q^\prime $
couplings with given $ v_S $:
\[
| ( \lambda_Q )_{ab} | , \ | ( \lambda_Q^\prime )_{ab} |
\sim \frac{m_Q}{v_S}
\]
Then, the $ f_Q $, $ f_Q^\prime $ and $ h_q $ couplings
are taken so as to reproduce the expected values of
the $ q $-$ Q $ mixing parameters:
\[
| ( f_Q )_{ia} | , \ | ( f_Q^\prime )_{ia} |
\rightarrow \epsilon^f_i \lesssim 1 \ , \
| ( h_q )_{ai} | \rightarrow \epsilon^h_i \lesssim 1 \ .
\]
The complex phases of these Yukawa couplings
and the VEV of the singlet Higgs field are taken randomly
in the full range:
\[
\arg [ h_q , f_Q , f_Q^\prime , \lambda_Q , \lambda_Q^\prime ] , \
\phi_S \in [ - \pi , \pi ] \ .
\]
The actual masses of the ordinary quarks are reproduced
by adjusting the relevant parameters as
\[
\begin{array}{l}
{\mbox{cases (a) and (b)}} :
\lambda_{q_i} \rightarrow m_{q_i} ,
\\
{\mbox{seesaw}} :
\left\{
\begin{array}{l}
( f_Q )_{ia} , ( h_q )_{ai} \rightarrow m_{q_i} \
( m_{Q_a} \sim m_Q ) \ ,
\\
m_{Q_i} \rightarrow m_{q_i} \
( m_{Q_1} \gg m_{Q_2} \gg m_{Q_3} ) \ .
\end{array} \right.
\end{array}
\]

\begin{figure}
\begin{center}
\hspace*{-0.8cm}
\includegraphics[height=10cm]{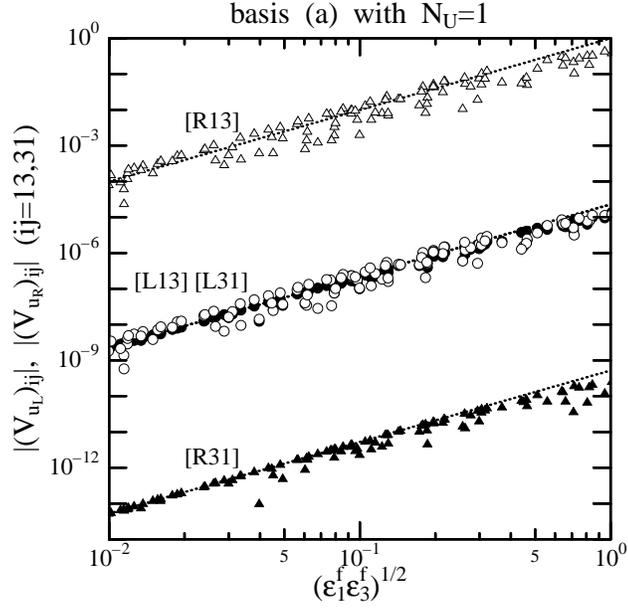}
\end{center}
\caption{
The $ u $-$ t $ mixing elements are shown
for the basis (a) with $ N_U = 1 $
depending on $ ( \epsilon^f_1 \epsilon^f_3 )^{1/2} $.
The values of the relevant couplings are taken randomly.
The marks are assigned as
``circle" : $ | ( V_{u_{\rm L}} )_{ij} | $
and ``triangle" : $ | ( V_{u_{\rm R}} )_{ij} | $.
The respective elements are denoted as
``blank mark" : $ ij = 13 $ and ``filled mark" : $ ij = 31 $.
The dotted lines indicate the expected flavor structures.
\label{f1}}
\end{figure}

\begin{figure}
\begin{center}
\hspace*{-0.8cm}
\includegraphics[height=10cm]{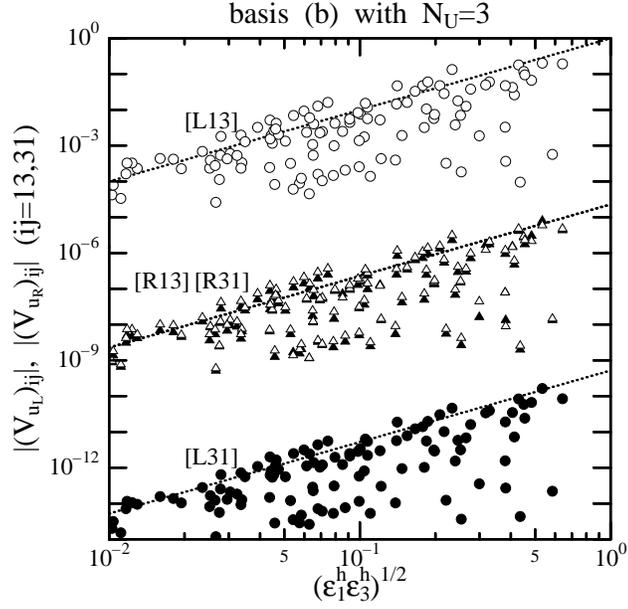}
\end{center}
\caption{
The $ u $-$ t $ mixing elements are shown
for the basis (b) with $ N_U = 3 $
depending on $ ( \epsilon^h_1 \epsilon^h_3 )^{1/2} $.
The marks are assigned the same as in Fig. \ref{f1}.
The dotted lines indicate the expected flavor structures.
\label{f2}}
\end{figure}

\begin{figure}
\begin{center}
\hspace*{-0.8cm}
\includegraphics[height=10cm]{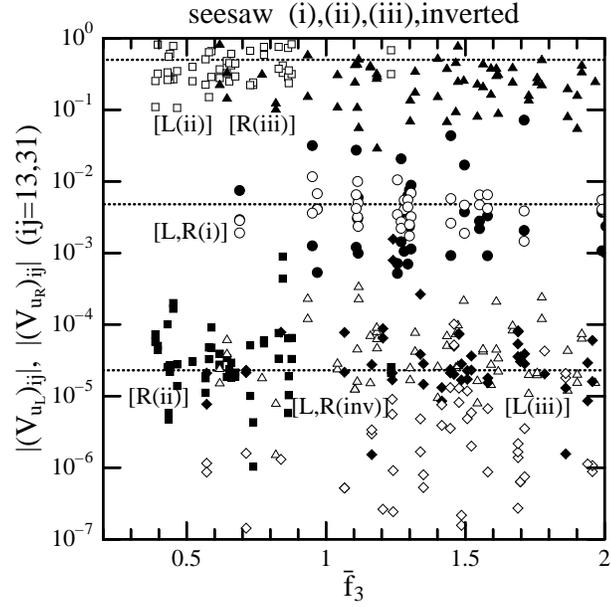}
\end{center}
\caption{
The $ u $-$ t $ mixing elements are shown depending on $ {\bar f}_3 $
for the seesaw models of (i), (ii), (iii) and inverted cases.
The marks are assigned as
``circle" : (i), ``square" : (ii), ``triangle" : (iii)
and ``diamond" : inverted.
The chirality of the mixings is also denoted as
``blank mark" : $ | ( V_{u_{\rm L}} )_{13,31} | $
and ``filled mark" : $ | ( V_{u_{\rm R}} )_{13,31} | $.
The dotted lines indicate the expected values.
\label{f3}}
\end{figure}

\begin{figure}
\begin{center}
\hspace*{-0.8cm}
\includegraphics[height=10cm]{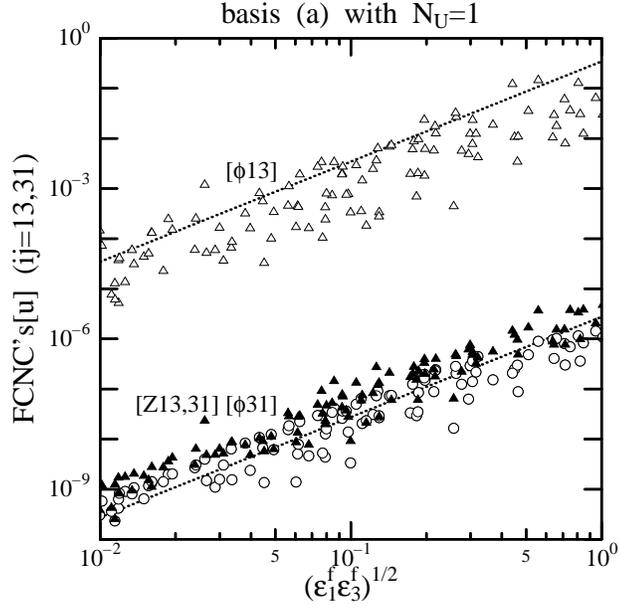}
\end{center}
\caption{
The FCNC's of the $ u $-$ t $ transition are shown
for the basis (a) with $ N_U = 1 $
depending on $ ( \epsilon^f_1 \epsilon^f_3 )^{1/2} $.
The marks are assigned as
``circle (blank)": $ | \Delta {\cal Z}_{\cal U} [u]_{13} |
= | \Delta {\cal Z}_{\cal U} [u]_{31} | $,
``triangle (blank)" : $ {\bar \Lambda}_{\cal U} [u]_{13} $
and ``triangle (filled)" : $ {\bar \Lambda}_{\cal U} [u]_{31} $.
The dotted lines indicate the expected flavor structures.
\label{f4}}
\end{figure}

\begin{figure}
\begin{center}
\hspace*{-0.8cm}
\includegraphics[height=10cm]{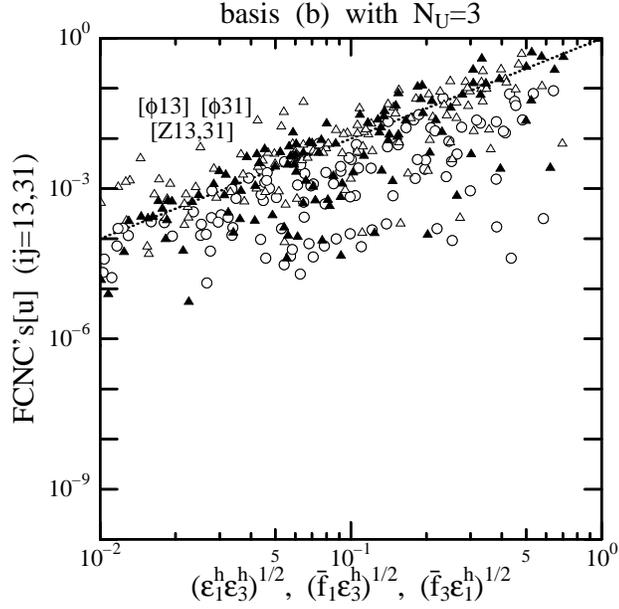}
\end{center}
\caption{
The FCNC's of the $ u $-$ t $ transition are shown
for the basis (b) with $ N_U = 3 $.
They depend, respectively, on $ ( \epsilon^h_1 \epsilon^h_3 )^{1/2} $
for the gauge couplings
and $ ( {\bar f}_i \epsilon^h_j )^{1/2} $ ($ ij = 13, 31 $)
for the scalar couplings.
The marks are assigned the same as in Fig. \ref{f4}.
The dotted lines indicate the expected flavor structures.
\label{f5}}
\end{figure}

\begin{figure}
\begin{center}
\hspace*{-0.8cm}
\includegraphics[height=10cm]{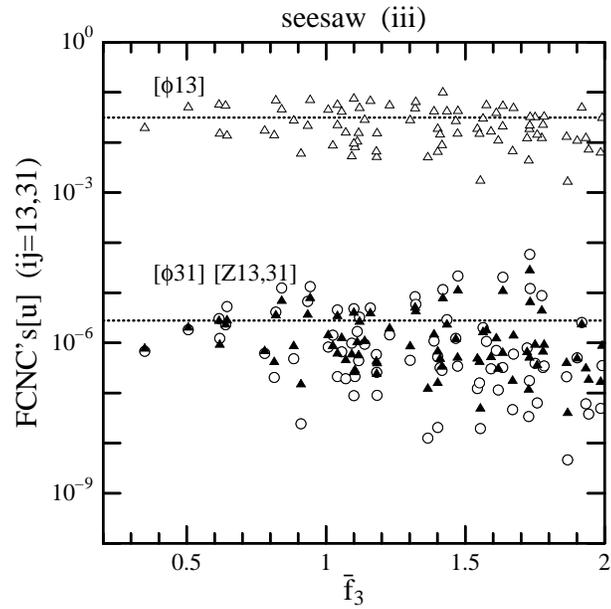}
\end{center}
\caption{
The FCNC's of the $ u $-$ t $ transition are shown
for the seesaw model of (iii) depending on $ {\bar f}_3 $.
The dotted lines indicate the expected values.
\label{f6}}
\end{figure}

\begin{figure}
\begin{center}
\hspace*{-0.8cm}
\includegraphics[height=10cm]{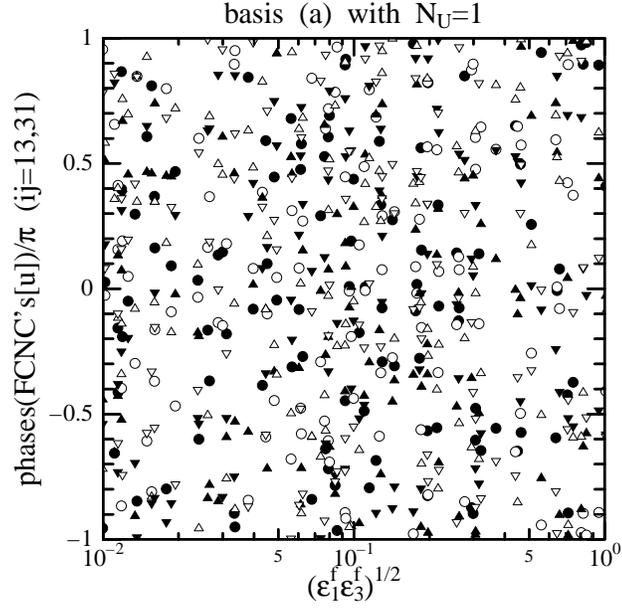}
\end{center}
\caption{
The complex phases involved in the FCNC's of the $ u $-$ t $ transition
are shown for the basis (a) with $ N_U = 1 $.
The marks are assigned as
``circle" : $ \arg [ \Delta {\cal Z}_{\cal U} [u]_{ij} ] $,
``triangle-up" : $ \arg [ \Lambda_{\cal U}^{S_+} [u]_{ij} ] $
and ``triangle-down" : $ \arg [ \Lambda_{\cal U}^{S_-} [u]_{ij} ] $.
The respective elements are denoted as
``blank mark" : $ ij = 13 $ and ``filled mark" : $ ij = 31 $.
\label{f7}}
\end{figure}

The quark mass matrix $ {\cal M}_{\cal Q} $
given with these parameters is diagonalized numerically.
Then, the quark mixing matrices are determined precisely,
and the flavor changing couplings are calculated.
Some typical results on these $ q $-$ Q $ mixing effects are
shown in Figs. \ref{f1} -- \ref{f7}
by noting for instance the $ u $-$ t $ transition terms.
Similar results are also obtained for the other flavor changing terms.

In Fig. \ref{f1}, the $ u $-$ t $ mixing elements
are shown for the basis (a) with $ N_U = 1 $
depending on the combination $ ( \epsilon^f_1 \epsilon^f_3 )^{1/2} $
of the $ q $-$ Q $ mixing parameters.
The values of the relevant couplings are taken randomly.
The marks are assigned as
\[
\begin{array}{lcl}
{\rm circle} & : & | ( V_{u_{\rm L}} )_{ij} | \ {\mbox{(left-handed)}} , \\
{\rm triangle} & : & | ( V_{u_{\rm R}} )_{ij} | \ {\mbox{(right-handed)}} .
\end{array}
\]
The respective elements are denoted as
\[
\begin{array}{lcl}
{\mbox{blank mark}} & : & ij = 13 , \\
{\mbox{filled mark}} & : & ij = 31 .
\end{array}
\]
The dotted lines indicate the expected flavor structures
given in Eqs. (\ref{eqn:VqL-a}) and (\ref{eqn:VqR-a}).
In Fig. \ref{f2}, the same quantities as in Fig. \ref{f1}
are shown for the basis (b) with $ N_U = 3 $.
The dotted lines indicate the expected flavor structures
given in Eqs. (\ref{eqn:VqL-b}) and (\ref{eqn:VqR-b}).
The duality (\ref{eqn:a-b}) between the bases (a) and (b)
is clearly observed in these figures.
That is, the regions of the circles (left-handed mixing)
and the triangles (right-handed mixing) are exchanged.
It is also found that similar $ q $-$ Q $ mixing effects are obtained
irrespectively of the number $ N_Q $ of the singlet quarks.

In Fig. \ref{f3}, the $ u $-$ t $ mixing elements
are shown for the seesaw models of (i), (ii), (iii) and inverted
cases depending on the relevant coupling parameter
$ {\bar f}_3 \equiv \sum_a  | ( f_U )_{3a} | / 3 $.
It should here be remarked that $ m_t \sim v $ is obtained
for $ | ( f_U )_{3a} | \sim 1 $ and $ | ( h_u )_{a3} | \sim 1 $
with $ m_Q \sim v_S $.
The marks are assigned for the respective cases as
\[
\begin{array}{lcl}
{\rm circle} & : & {\mbox{(i)}} , \\
{\rm square} & : & {\mbox{(ii)}} , \\
{\rm triangle} & : & {\mbox{(iii)}} , \\
{\rm diamond} & : & {\mbox{inverted}} .
\end{array}
\]
The chirality of the mixings is also denoted as
\[
\begin{array}{lcl}
{\mbox{blank mark}} & : & | ( V_{u_{\rm L}} )_{13, 31} | \
{\mbox{(left-handed)}} , \\
{\mbox{filled mark}} & : & | ( V_{u_{\rm R}} )_{13, 31} | \
{\mbox{(right-handed)}} .
\end{array}
\]
Here, $ | ( V_{u_\chi} )_{13} | $ and $ | ( V_{u_\chi} )_{31} | $
appear to be of the same order in the seesaw models.
The dotted lines indicate the expected values
given in Eqs. (\ref{eqn:qQmix-i}), (\ref{eqn:qQmix-ii}),
(\ref{eqn:qQmix-iii}), (\ref{eqn:VqL-ssi}) and (\ref{eqn:VqR-ssi}).
We observe, in particular, that the significant mixings are induced
for $ ( V_{u_{\rm L}} )_{13} $ (blank square) of case (ii)
and $ ( V_{u_{\rm R}} )_{31} $ (filled triangle) of case (iii).
Although the left-handed mixings $ ( V_{q_{\rm L}} )_{ij} $
appear to be of $ O(1) $ for the case (ii), its unitarity violation
is suppressed by $ \epsilon^h_i \epsilon^h_j \sim ( {\bar h} v / m_Q )^2 $.
Hence, in this specific case with $ m_t \sim {\bar h} v $,
the singlet quark masses $ \sim m_U $ will be required
to be sufficiently larger than the electroweak scale
\cite{LL,qQmix}.
The right-handed mixings $ ( V_{q_{\rm R}} )_{ij} \sim 1 $
for the case (iii), on the other hand, contribute
to provide the significant scalar couplings
$ \Lambda_{\cal Q}^\alpha [q]_{ij}
\sim ( m_{q_j} / v_S )( {\bar f}_3 v_S / m_Q )^2 $.

The magnitudes of the FCNC's of the $ u $-$ t $ transition
are shown in Figs. \ref{f4}, \ref{f5} and \ref{f6}, respectively,
for the basis (a), basis (b) and seesaw of case (iii).
In these cases, the significant FCNC's ($ \phi $)
are obtained, which are not related to the FCNC's ($ Z $).
The marks are assigned as
\[
\begin{array}{lcl}
{\mbox{circle (blank)}} & : & | \Delta {\cal Z}_{\cal U} [u]_{13} |
= | \Delta {\cal Z}_{\cal U} [u]_{31} | , \\
{\mbox{triangle (blank)}} & : & {\bar \Lambda}_{\cal U} [u]_{13} , \\
{\mbox{triangle (filled)}} & : & {\bar \Lambda}_{\cal U} [u]_{31} ,
\end{array}
\]
where
\[
{\bar \Lambda}_{\cal Q} [q]_{ij}
\equiv \frac{1}{3} \left[ | \Lambda_{\cal Q}^H [q]_{ij} |
+ | \Lambda_{\cal Q}^{S_+} [q]_{ij} |
+ | \Lambda_{\cal Q}^{S_-} [q]_{ij} | \right] .
\]
These FCNC's are shown depending on the relevant parameters
$ \epsilon^f_i $, $ \epsilon^h_i $
and $ {\bar f}_i \equiv \sum_a | ( f_U )_{ia} | / 3 $
($ | ( f_U )_{ia} | = | ( f_U^\prime )_{ia} | $
in the basis (b) with $ \Delta_{qQ}^\prime = {\bf 0} $),
so that their flavor structures are readily compared
to the expected ones (dotted lines).
We observe clearly that in the basis (a) and seesaw of case (iii)
the scalar coupling (13 element) of the $ u $-$ t $ transition
is quite significant being proportional to the top quark mass $ m_t $.
On the other hand, in the basis (b) the gauge couplings
as well as the scalar couplings can be considerable.

In Fig. \ref{f7}, the complex phases involved in the FCNC's
of the $ u $-$ t $ transition are shown for the basis (a).
The marks (blank for $ ij = 13 $ and filled for $ ij = 31 $)
are assigned as
\[
\begin{array}{lcl}
{\mbox{circle}} & : & \arg [ \Delta {\cal Z}_{\cal U} [u]_{ij} ] , \\
{\mbox{triangle-up}} & : & \arg [ \Lambda_{\cal U}^{S_+} [u]_{ij} ] , \\
{\mbox{triangle-down}} & : & \arg [ \Lambda_{\cal U}^{S_-} [u]_{ij} ] .
\end{array}
\]
[Note that $ \arg [ \Lambda_{\cal Q}^H [q]_{ij} ]
= \arg [ \Delta {\cal Z}_{\cal Q} [q]_{ij} ] $
due to Eq. (\ref{eqn:lmqhqtld-DZQ}).]
Here, the complex phases of the relevant parameters are taken randomly.
It is clearly observed that the $ CP $ violating phases
in the gauge and scalar couplings take various values
depending on the phases of the original model parameters.

\section{Summary and discussion}
\label{sec:summary}

The singlet quarks may provide various intriguing effects
in particle physics and cosmology
through the mixing with the ordinary quarks.
We have presented the systematic and comprehensive investigations
on the quark mixings in the electroweak models with singlet quarks.
There are some appropriate choices of the quark basis
for the electroweak eigenstates,
where the entire quark mass matrix $ {\cal M}_{\cal Q} $
has the specific form without loss of generality.
They are the basis (a) with the $ Q^c q $ mixing term
$ \Delta_{qQ}^\prime = {\bf 0} $,
the basis (b) with the $ q^c Q $ mixing term $ \Delta_{qQ} = {\bf 0} $
and the seesaw with the ordinary quark mass matrix $ M_q = {\bf 0} $.
We may take complementarily these quark bases,
depending on the model parameter ranges
and also the symmetries and matter contents.

We have examined in detail for these bases
how the ordinary quark masses and mixings are affected
by the $ q $-$ Q $ mixing.
The flavor changing interactions are also modified
by the $ q $-$ Q $ mixing.  Specifically, the CKM unitarity
within the ordinary quark sector is violated,
and the FCNC's arise both in the gauge and scalar couplings.
The structures of these flavor changing interactions
have been clarified for the respective quark bases.
In fact, they are described appropriately
in terms of the $ q $-$ Q $ mixing parameters and the quark masses.
These results ensure that there are some reasonable ranges
of the model parameters where the ordinary quark mass hierarchy
and the actual CKM structure are reproduced
even in the presence of singlet quarks.
In these meanings, the present case with singlet quarks
may provide an interesting extension
of the idea of natural flavor conservation
\cite{GWP}.

A detailed numerical analysis has further been performed
for calculating precisely the quark mixings and flavor changing couplings
with singlet quarks.
Then, it has been confirmed that the $ q $-$ Q $ mixing effects
really exhibit the expected flavor structures.
These calculations on the singlet quarks may be extended readily
for the models with various exotic quarks and leptons
such as vector-like electroweak doubles,
where the entire fermion mass matrix has the same form
as $ {\cal M}_{\cal Q} $.

We finally discuss some phenomenological implications
derived from the results of the present investigations.
We particularly note that the scalar FCNC's ($ \phi $) are sometimes
fairly larger than the gauge FCNC's ($ Z $),
as seen in Figs. \ref{f4}, \ref{f5} and \ref{f6}.
Then, if the singlet quarks and extra Higgs particles
exist just above the electroweak scale
with $ m_Q , m_{\phi_\alpha} \sim 100 {\rm GeV} - 1 {\rm TeV} $,
the scalar FCNC's ($ \phi $) are expected to provide significant effects
on the flavor changing and $ CP $ violating processes, e.g.,
the $ K $, $ B $ and $ D $ meson physics.
Hence, these effects of scalar FCNC's ($ \phi $)
rather than those of gauge FCNC's ($ Z $)
might serve as signals for the new physics beyond the standard model.
This possibility has not been paid so much attention before
in the models with singlet quarks.

The FCNC's induced by the $ q $-$ Q $ mixing
may sometimes be related to the ordinary quark masses.
Then, it will be expected that the $ q $-$ Q $ mixing effects
can be observed most likely in the top quark physics.
In particular, in the basis (a)
the scalar FCNC's ($ \phi $) and the gauge FCNC's ($ Z $)
arise, respectively, at the first and second orders
of the ordinary quark masses.
Then, we expect $ {\rm Br} ( t \rightarrow c Z ) \sim 10^{-5} $
with $ | ( \Delta {\cal Z}_{\cal U} )_{13} |
\sim ( m_c / m_Q )( m_t / m_Q ) $
for $ m_Q \sim 200 {\rm GeV} $ and $ \epsilon^f_i \sim 1 $
while negligibly small $ {\rm Br} ( t \rightarrow u Z ) $.
If the neutral Higgs particle $ \phi_0 $, which is mainly
$ h_1 $ of the standard model, is light enough,
we would obtain the top quark decays involving $ \phi_0 $
\cite{top-Higgs}.
The branching ratio can be rather significant
with the scalar couplings $ | ( \Lambda_{\cal Q}^{S_\pm} )_{i3} |
\sim ( m_t / v_S ) \epsilon^f_i \epsilon^f_3 $
and the $ H^0 $-$ S $ mixings $ O_{10} , O_{20} \sim v / v_S $.
In fact, we estimate $ {\rm Br} ( t \rightarrow q_i + \phi_0 )
\sim 10^{-2} ( \epsilon^f_i \epsilon^f_3 )^2 $ ($ q_i = u , c $)
for $ v_S \sim 500 {\rm GeV} $ and $ m_{\phi_0} \simeq 100{\rm GeV} $.

It will be worth making further investigations
of these remarkable effects particularly of the scalar FCNC's
induced by the $ q $-$ Q $ mixing.

\acknowledgments

We would like to thank R. Kitano and M. Senami for valuable discussions.

\appendix

\section{Diagonalization of the quark mass matrix}
\label{sec:diagonalization}

We here present the algebraic calculations
for diagonalizing the quark mass matrix $ {\cal M}_{\cal Q} $.
The leading order results for $ N_Q = 1 $ are given in the literature
\cite{AgBr}.
We do not intend to calculate explicitly the higher order corrections.
The following treatments hence seems to be of little practical use.
The diagonalization of $ {\cal M}_{\cal Q} $
can anyway be made precisely by numerical calculations.
We would rather like to present comprehensive explanations
for the specific flavor structures of the $ q $-$ Q $ mixing effects.
They are even valid beyond the leading orders for more general cases
with several singlet quarks and wide ranges of the model parameters.
Then, the following arguments appear to be helpful
to understand the results of precise numerical calculations.

The diagonalization of $ {\cal M}_{\cal Q} $ may be performed at two steps
by dividing the unitary transformation as
\begin{equation}
{\cal V}_{{\cal Q}_\chi}
= {\cal V}_{{\cal Q}_\chi}^{(1)} {\cal V}_{{\cal Q}_\chi}^{(2)}
\end{equation}
with
\begin{equation}
{\cal V}_{{\cal Q}_\chi}^{(1)}
= \left( \begin{array}{cc}
V_{q_\chi}^{(1)} & \epsilon_{q_\chi}^{(1)} \\
- \epsilon_{q_\chi}^{(1) \dagger} & V_{Q_\chi}^{(1)}
\end{array} \right) ,
\end{equation}
\begin{equation}
{\cal V}_{{\cal Q}_\chi}^{(2)}
= \left( \begin{array}{cc}
V_{q_\chi}^{(2)} & {\bf 0} \\
{\bf 0} & V_{Q_\chi}^{(2)} \end{array} \right) .
\end{equation}
Here, $ V_{q_\chi}^{(1)} $ and $ V_{Q_\chi}^{(1)} $
are in general non-unitary
due to the $ q $-$ Q $ mixing $ \epsilon_{q_\chi}^{(1)} $
($ \epsilon_{q_\chi}^{(1) \prime} = \epsilon_{q_\chi}^{(1)} $,
as given explicitly below),
while $ V_{q_\chi}^{(2)} $ and $ V_{Q_\chi}^{(2)} $
are unitary by definition with
$ \epsilon_{q_\chi}^{(2)} = \epsilon_{q_\chi}^{(2) \prime} = {\bf 0} $.
The components of the entire transformation $ {\cal V}_{{\cal Q}_\chi} $
are then given by
\begin{eqnarray}
\epsilon_{q_\chi} &=& \epsilon_{q_\chi}^{(1)} V_{Q_\chi}^{(2)} , \
\epsilon_{q_\chi}^\prime
= V_{q_\chi}^{(2) \dagger} \epsilon_{q_\chi}^{(1)} ,
\label{eqn:eqeqpchi} \\
V_{q_\chi} &=& V_{q_\chi}^{(1)} V_{q_\chi}^{(2)} , \
V_{Q_\chi} = V_{Q_\chi}^{(1)} V_{Q_\chi}^{(2)} .
\label{eqn:VqVQchi}
\end{eqnarray}

The first step transformation is utilized
for eliminating $ \Delta_{qQ} $ and $ \Delta_{qQ}^\prime $,
which may be given by
\begin{equation}
{\cal V}_{{\cal Q}_\chi}^{(1)} = \exp[ {\cal E}_{{\cal Q}_\chi} ]
= {\bf 1} + {\cal E}_{{\cal Q}_\chi}
+ \frac{1}{2!} {\cal E}_{{\cal Q}_\chi}^2 + \cdots
\end{equation}
with certain anti-hermitian matrix
\begin{equation}
{\cal E}_{{\cal Q}_\chi} = \left( \begin{array}{cc}
{\bf 0} & {\bar \epsilon}_{q_\chi} \\
- {\bar \epsilon}^\dagger_{q_\chi} & {\bf 0}
\end{array} \right) .
\end{equation}
Then, we have the submatrices in $ {\cal V}_{{\cal Q}_\chi}^{(1)} $ as
\begin{eqnarray}
\epsilon_{q_\chi}^{(1)} &=& \epsilon_{q_\chi}^{(1) \prime}
= {\bar \epsilon}_{q_\chi}
- \frac{1}{3!} {\bar \epsilon}_{q_\chi} {\bar \epsilon}_{q_\chi}^\dagger
{\bar \epsilon}_{q_\chi} + \ldots
\nonumber \\
&=& {\bar \epsilon}_{q_\chi} {\bar B}_{q_\chi}
= {\bar B}_{q_\chi}^\prime {\bar \epsilon}_{q_\chi} ,
\label{eqn:eqeqp1}
\\
V_{q_\chi}^{(1)} &=& V_{q_\chi}^{(1) \dagger} = {\bf 1}
- \frac{1}{2!}{\bar \epsilon}_{q_\chi}{\bar \epsilon}_{q_\chi}^\dagger
+ \cdots
\nonumber \\
&=& {\bf 1}
+ {\bar \epsilon}_{q_\chi} {\bar A}_{q_\chi} {\bar \epsilon}_{q_\chi}^\dagger
= {\bf 1}
+ \epsilon_{q_\chi}^{(1)} A_{q_\chi} \epsilon_{q_\chi}^{(1) \dagger} ,
\label{eqn:Vq1}
\\
V_{Q_\chi}^{(1)} &=& V_{Q_\chi}^{(1) \dagger} = {\bf 1}
- \frac{1}{2!}{\bar \epsilon}_{q_\chi}^\dagger {\bar \epsilon}_{q_\chi}
+ \cdots
\nonumber \\
&=& {\bf 1}
+ {\bar \epsilon}_{q_\chi}^\dagger {\bar A}_{Q_\chi} {\bar \epsilon}_{q_\chi}
= {\bf 1}
+ \epsilon_{q_\chi}^{(1) \dagger} A_{Q_\chi} \epsilon_{q_\chi}^{(1)} ,
\label{eqn:VQ1}
\end{eqnarray}
where $ {\bar A}_{q_\chi} $, $ {\bar A}_{Q_\chi} $, $ {\bar B}_{q_\chi} $,
$ {\bar B}_{q_\chi}^\prime $
$ = {\bf 1} + O( {\bar \epsilon}_{q_\chi}^2 ) $,
and $ A_{q_\chi} = {\bar B}_{q_\chi}^{-1}
{\bar A}_{q_\chi} {\bar B}_{q_\chi}^{\dagger -1} $,
$ A_{Q_\chi} = {\bar B}_{q_\chi}^{\prime \dagger -1}
{\bar A}_{Q_\chi} {\bar B}_{q_\chi}^{\prime -1} $.
The quark mass matrix is transformed as
\begin{equation}
{\cal V}_{{\cal Q}_{\rm R}}^{(1) \dagger}
{\cal M}_{\cal Q} {\cal V}_{{\cal Q}_{\rm L}}^{(1)}
= \left( \begin{array}{cc}
M_q^{(1)} & {\bf 0} \\ {\bf 0} & M_Q^{(1)}
\end{array} \right) .
\end{equation}
Here, the first step transformation is determined by the conditions,
\begin{eqnarray}
\Delta^{(1)}_{qQ}
&=& V_{q_{\rm R}}^{(1) \dagger} ( M_q \epsilon_{q_{\rm L}}^{(1)}
+ \Delta_{qQ} V_{Q_{\rm L}}^{(1)} )
\nonumber \\
&-& \epsilon_{q_{\rm R}}^{(1)} ( M_Q V_{Q_{\rm L}}^{(1)}
+ \Delta_{qQ}^\prime \epsilon_{q_{\rm L}}^{(1)} ) = {\bf 0} ,
\label{eqn:DqQ10} \\
\Delta^{(1) \prime}_{qQ}
&=& ( V_{Q_{\rm R}}^{(1) \dagger} \Delta_{qQ}^\prime
+ \epsilon_{q_{\rm R}}^{(1) \dagger} M_q ) V_{q_{\rm L}}^{(1)}
\nonumber \\
&-& ( V_{Q_{\rm R}}^{(1) \dagger} M_Q
+ \epsilon_{q_{\rm R}}^{(1) \dagger} \Delta_{qQ} )
\epsilon_{q_{\rm L}}^{(1) \dagger} = {\bf 0} .
\label{eqn:DqQp10}
\end{eqnarray}
These $ 2 \times 3 \times N_Q $ conditions are just satisfied
by the $ 2 \times 3 \times N_Q $ parameters contained
in $ {\bar \epsilon}_{q_{\rm L}} $ and $ {\bar \epsilon}_{q_{\rm R}} $.
Practically, in these conditions (\ref{eqn:DqQ10}) and (\ref{eqn:DqQp10})
the ordinary quark mixing matrices $ V_{q_\chi}^{(1)} $
may be expressed in terms of the $ q $-$ Q $ mixing matrices
$ \epsilon_{q_\chi}^{(1)} $ with Eq. (\ref{eqn:Vq1}).
Then, by performing some algebra we obtain the relations,
\begin{eqnarray}
\epsilon_{q_{\rm L}}^{(1)}
&=& ( \Delta_{qQ}^{\prime \dagger} + 
M_q^\dagger \epsilon_{q_{\rm R}}^{(1)} V_{Q_{\rm R}}^{(1) -1} )
( M_Q^\dagger + \delta_{Q_{\rm L}}^\dagger )^{-1} ,
\label{eqn:eqL1}
\\
\epsilon_{q_{\rm R}}^{(1)}
&=& ( \Delta_{qQ} + M_q \epsilon_{q_{\rm L}}^{(1)} V_{Q_{\rm L}}^{(1) -1} )
( M_Q + \delta_{Q_{\rm R}} )^{-1} ,
\label{eqn:eqR1}
\end{eqnarray}
where
\begin{eqnarray}
\delta_{Q_{\rm L}}
&=& V_{Q_{\rm R}}^{(1) \dagger -1}
\epsilon_{q_{\rm R}}^{(1) \dagger} \Delta_{qQ}
- \Delta_{qQ}^\prime \epsilon_{q_{\rm L}}^{(1)} A_{q_{\rm L}}^\dagger
\nonumber \\
&-& V_{Q_{\rm R}}^{(1) \dagger -1} \epsilon_{q_{\rm R}}^{(1) \dagger}
M_q \epsilon_{q_{\rm L}}^{(1)} A_{q_{\rm L}}^\dagger ,
\\
\delta_{Q_{\rm R}}
&=& - A_{q_{\rm R}} \epsilon_{q_{\rm R}}^{(1) \dagger} \Delta_{qQ}
+ \Delta_{qQ}^{\prime \dagger}
\epsilon_{q_{\rm L}}^{(1)} V_{Q_{\rm L}}^{(1) -1}
\nonumber \\
&-& A_{q_{\rm R}} \epsilon_{q_{\rm R}}^{(1) \dagger}
M_q \epsilon_{q_{\rm L}}^{(1)} V_{Q_{\rm L}}^{(1) -1} .
\end{eqnarray}

The effective quark mass matrices obtained at the first step are given by
\begin{eqnarray}
M_q^{(1)} &=& ( V_{q_{\rm R}}^{(1) \dagger} M_q
- \epsilon_{q_{\rm R}}^{(1)} \Delta_{qQ}^\prime )
V_{q_{\rm L}}^{(1)}
\nonumber \\
&-& ( V_{q_{\rm R}}^{(1) \dagger} \Delta_{qQ}
- \epsilon_{q_{\rm R}}^{(1)} M_Q ) \epsilon_{q_{\rm L}}^{(1) \dagger} ,
\\
M_Q^{(1)} &=& ( V_{Q_{\rm R}}^{(1) \dagger} M_Q
+ \epsilon_{q_{\rm R}}^{(1) \dagger} \Delta_{qQ} ) V_{Q_{\rm L}}^{(1)}
\nonumber \\
&+& ( \epsilon_{q_{\rm R}}^{(1) \dagger} M_q
+ V_{Q_{\rm R}}^{(1) \dagger} \Delta_{qQ}^\prime )
\epsilon_{q_{\rm L}}^{(1)} .
\end{eqnarray}
Here, $ M_q^{(1)} $ for the ordinary quarks is, in particular,
calculated by considering the relation
$ - ( V_{q_{\rm R}}^{(1) \dagger} \Delta_{qQ}
- \epsilon_{q_{\rm R}}^{(1)} M_Q )
= ( V_{q_{\rm R}}^{(1) \dagger} M_q
- \epsilon_{q_{\rm R}}^{(1)} \Delta_{qQ}^\prime )
\epsilon_{q_{\rm L}}^{(1)} V_{Q_{\rm L}}^{(1) -1} $
from Eq. (\ref{eqn:DqQ10}) as
\begin{equation}
M_q^{(1)} = ( V_{q_{\rm R}}^{(1) \dagger} M_q
- \epsilon_{q_{\rm R}}^{(1)} \Delta_{qQ}^\prime ) ( {\bf 1} + R_q ) \ ,
\label{eqn:Mq1}
\end{equation}
where
\begin{equation}
R_q = \epsilon_{q_{\rm L}}^{(1)}
( A_{q_{\rm L}} + V_{Q_{\rm L}}^{(1) -1} )
\epsilon_{q_{\rm L}}^{(1) \dagger} .
\label{eqn:Rq1}
\end{equation}

The effective mass matrices $ M_q^{(1)} $ and $ M_Q^{(1)} $
at the first step are generally non-diagonal.
They are diagonalized at the second step as
\begin{eqnarray}
V_{q_{\rm R}}^{(2) \dagger} M_q^{(1)} V_{q_{\rm L}}^{(2)}
&=& {\bar M}_q ,
\\
V_{Q_{\rm R}}^{(2) \dagger} M_Q^{(1)} V_{Q_{\rm L}}^{(2)}
&=& {\bar M}_Q .
\end{eqnarray}
This completes the diagonalization of $ {\cal M}_{\cal Q} $.

\subsection{Basis (a) with $ \Delta_{qQ}^\prime = {\bf 0} $}

In the basis (a) with $ \Delta_{qQ}^\prime = {\bf 0} $,
the $ q $-$ Q $ mixing is generated by the $ \Delta_{qQ} $ term.
Then, the $ q $-$ Q $ mixing matrices at the first step
are determined with Eqs. (\ref{eqn:eqL1}) and (\ref{eqn:eqR1}) as
\begin{eqnarray}
( \epsilon^{(1)}_{q_{\rm L}} )_{ia}
& \sim & ( m_{q_i} / m_Q ) \epsilon^f_i ,
\label{eqn:eqL1-a}
\\
( \epsilon^{(1)}_{q_{\rm R}} )_{ia}
& \sim & \epsilon^f_i ,
\label{eqn:eqR1-a}
\end{eqnarray}
where the relation $ m_{q_i}^0 \sim m_{q_i} $,
as seen later in Eq. (\ref{eqn:mqi-a}),
is considered for $ ( M_q )_{ij} = m_{q_i}^0 \delta_{ij} $.
We also obtain form Eq. (\ref{eqn:Vq1})
with Eqs. (\ref{eqn:eqL1-a}) and (\ref{eqn:eqR1-a})
\begin{eqnarray}
( V_{q_{\rm L}}^{(1)} )_{ij} & \sim & \delta_{ij}
+ ( m_{q_i} / m_Q ) ( m_{q_j} / m_Q ) \epsilon^f_i \epsilon^f_j ,
\\
( V_{q_{\rm R}}^{(1)} )_{ij} & \sim & \delta_{ij}
+ \epsilon^f_i \epsilon^f_j .
\end{eqnarray}

The effective mass matrix $ M_q^{(1)} $ for the ordinary quarks is given
from Eq. (\ref{eqn:Mq1}) with $ \Delta_{qQ}^\prime = {\bf 0} $ as
\begin{equation}
M_q^{(1)} = V_{q_{\rm R}}^{(1) \dagger} M_q
( {\bf 1} + R_q ) .
\label{eqn:Mq1-a}
\end{equation}
The structure of $ R_q $ is specified in Eq. (\ref{eqn:Rq1})
with Eq. (\ref{eqn:eqL1-a}) as
\begin{equation}
( R_q )_{ij}
\sim ( m_{q_i} / m_Q ) ( m_{q_j} / m_Q ) \epsilon^f_i \epsilon^f_j .
\label{eqn:Rq1-a}
\end{equation}
It is suitable to modify $ M_q^{(1)} $ as
\begin{equation}
M_q^{(1^\prime)}
= V_{q_{\rm R}}^{( 2^\prime ) \dagger} M_q^{(1)}
= V_{q_{\rm R}}^{( 1^\prime ) \dagger} M_q ( {\bf 1} + R_q )
\end{equation}
by deforming $ V_{q_{\rm R}}^{(1)} $ into a triangular form
\begin{equation}
V_{q_{\rm R}}^{( 1^\prime )}
= V_{q_{\rm R}}^{(1)} V_{q_{\rm R}}^{( 2^\prime )}
\sim
\left( \begin{array}{ccc}
1 & \epsilon^f_1 \epsilon^f_2 & \epsilon^f_1 \epsilon^f_3 \\
0 & 1 & \epsilon^f_2 \epsilon^f_3 \\
0 & 0 & 1 \end{array} \right) \ .
\label{eqn:VqR1p-a}
\end{equation}
Here, the relevant unitary transformation is given as
\begin{equation}
( V_{q_{\rm R}}^{( 2^\prime )} )_{ij}
\sim \delta_{ij} + \epsilon^f_i \epsilon^f_j .
\end{equation}
Then, by considering the hierarchy
$ m_{q_1} \ll m_{q_2} \ll m_{q_3} \lesssim m_Q $, we obtain the relations,
\begin{eqnarray}
( M_q^{(1^\prime) \dagger} M_q^{(1^\prime)} )_{ij}
& \sim & \delta_{ij} m_{q_i}^2
+ m_{q_i} m_{q_j} \epsilon^f_i \epsilon^f_j ,
\label{eqn:Mq1p-1} \\
( M_q^{(1^\prime)} M_q^{(1^\prime) \dagger} )_{ij}
& \sim & \delta_{ij} m_{q_i}^2
+ \theta^k_{ij} m_{q_k}^2 \epsilon^f_i \epsilon^f_j ,
\label{eqn:Mq1p-2}
\end{eqnarray}
where
\[
\theta^k_{ij} = \left\{ \begin{array}{ll}
1 & {\mbox{($ k = i $ for $ i < j $, $ k = j $ for $ i > j $)}} \\
0 & {\mbox{(otherwise)}} \end{array} \right. .
\]

The effective mass matrix $ M_q^{(1^\prime)} $
and its squared ones (\ref{eqn:Mq1p-1}) and (\ref{eqn:Mq1p-2})
are diagonalized by the unitary transformations,
\begin{eqnarray}
( V_{q_{\rm L}}^{(2)} )_{ij} & \sim & \delta_{ij}
+ \frac{m_{q_i} m_{q_j}}{m_{q_i}^2 + m_{q_j}^2}
\epsilon^f_i \epsilon^f_j ,
\label{eqn:VqL2-a} \\
( V_{q_{\rm R}}^{( 2^{\prime \prime} )} )_{ij} & \sim & \delta_{ij}
+ \frac{m_{q_j}^2}{m_{q_i}^2 + m_{q_j}^2}
\epsilon^f_i \epsilon^f_j .
\label{eqn:VqR2pp-a}
\end{eqnarray}
The second step transformation of the right-handed ordinary quarks
is then given by
\begin{equation}
V_{q_{\rm R}}^{(2)}
= V_{q_{\rm R}}^{( 2^\prime )} V_{q_{\rm R}}^{( 2^{\prime \prime} )} .
\label{eqn:VqR2-a}
\end{equation}
The ordinary quark masses are obtained as
\begin{equation}
m_{q_i} = ( \lambda_{q_i} v / {\sqrt 2} )
\left[ 1 + \xi_{q_i} ( {\mbox{\boldmath $ \epsilon $}}^f )
( \epsilon^f_i )^2 \right]
\label{eqn:mqi-a}
\end{equation}
with certain factors
$ \xi_{q_i} ( {\mbox{\boldmath $ \epsilon $}}^f ) \sim 1 $,
which are in fact $ - 1/2 $ in the leading order for $ N_Q = 1 $.

The net effects on the quark mixings involving the ordinary quarks
are calculated in Eqs. (\ref{eqn:eqeqpchi}) and (\ref{eqn:VqVQchi}) as
\begin{eqnarray}
\epsilon_{q_\chi} &,& \epsilon_{q_\chi}^\prime
\sim \epsilon^{(1)}_{q_\chi} \ ( \chi = {\rm L} , {\rm R} ) , \\
\label{eqn:eqchi-a}
V_{q_{\rm L}} & \sim & V_{q_{\rm L}}^{(2)} , \
V_{q_{\rm R}} \sim V_{q_{\rm R}}^{(2^{\prime \prime})} .
\label{eqn:Vqchi-a}
\end{eqnarray}
The symbol ``$ \sim $" henceforth indicates
that the mixing matrices have the same structure
with respect to the ordinary quark flavors.
These relations are justified as follows.
The left-handed $ q $-$ Q $ mixing matrix $ \epsilon_{q_{\rm L}} $
is calculated as
$ ( \epsilon_{q_{\rm L}} )_{ia} = ( \epsilon_{q_{\rm L}}^{(1)} )_{ib}
( V_{Q_\chi}^{(2)} )_{ba} \sim ( m_{q_i} / m_Q ) \epsilon^f_i $
with $ ( V_{Q_\chi}^{(2)} )_{ba} \lesssim 1 $.
The calculation of $ \epsilon_{q_{\rm L}}^\prime $ is made as
\[
( \epsilon_{q_{\rm L}}^\prime )_{ia}
= ( V_{q_{\rm L}}^{(2) \dagger} )_{ii}
( \epsilon_{q_{\rm L}}^{(1)} )_{ia}
+ \sum_{j \not= i} ( V_{q_{\rm L}}^{(2) \dagger} )_{ij}
( \epsilon_{q_{\rm L}}^{(1)} )_{ja} .
\]
The first term is of the order of $ ( m_{q_i} / m_Q ) \epsilon^f_i $
for $ ( V_{q_{\rm L}}^{(2) \dagger} )_{ii} \sim 1 $.
The second term is estimated
as $ ( m_{q_i} / m_Q ) \epsilon^f_i ( \epsilon^f_j )^2 $
by using Eqs. (\ref{eqn:eqL1-a}) and (\ref{eqn:eqR1-a}) and the relation
\[
\frac{m_{q_i} m_{q_j}}{m_{q_i}^2 + m_{q_j}^2} m_{q_j} < m_{q_i} .
\]
It is in fact smaller than the first term.
Similar calculations are made for
the right-handed $ q $-$ Q $ mixing matrices
$ \epsilon_{q_{\rm R}} $ and $ \epsilon_{q_{\rm R}}^\prime $.
For the left-handed ordinary quark mixing,
the relation $ ( V_{q_{\rm L}} )_{ij} \sim ( V_{q_{\rm L}}^{(2)} )_{ij} $
is verified by considering
the inequality $ ( V_{q_{\rm L}}^{(1)} )_{ik} ( V_{q_{\rm L}}^{(2)} )_{kj}
\lesssim ( V_{q_{\rm L}}^{(1)} )_{ij}
\lesssim ( V_{q_{\rm L}}^{(2)} )_{ij} $
($ k \not= i,j $) for $ m_{q_i} \lesssim m_Q $.
The right-handed ordinary quark mixing is
estimated with the relation
\begin{equation}
V_{q_{\rm R}} = V_{q_{\rm R}}^{(1)} V_{q_{\rm R}}^{(2)}
= V_{q_{\rm R}}^{( 1^\prime )} V_{q_{\rm R}}^{( 2^{\prime \prime} )} ,
\end{equation}
where Eqs. (\ref{eqn:VqR1p-a}) and (\ref{eqn:VqR2-a}) are considered
in the second equality.  By using the specific forms
(\ref{eqn:VqR1p-a}) and (\ref{eqn:VqR2pp-a})
for $ V_{q_{\rm R}}^{( 1^\prime )} $
and $ V_{q_{\rm R}}^{( 2^{\prime \prime} )} $,
we find the relation $ ( V_{q_{\rm R}} )_{ij}
\sim ( V_{q_{\rm R}}^{( 2^{\prime \prime} )} )_{ij} $
under the quark mass hierarchy
$ m_{q_1} \ll m_{q_2} \ll m_{q_3} \lesssim m_Q $.

\subsection{Basis (b) with $ \Delta_{qQ} = {\bf 0} $}

The diagonalization of $ {\cal M}_{\cal Q} $
in the basis (b) with $ \Delta_{qQ} = {\bf 0} $ is practically
reduced to what is made in the basis (a)
with $ \Delta_{qQ}^\prime = {\bf 0} $.
The hermite conjugate of $ {\cal M}_{\cal Q} $
with $ \Delta_{qQ} = {\bf 0} $ is given by
\begin{equation}
{\cal M}_{\cal Q}^\dagger = \left( \begin{array}{cc}
M_q^\dagger & \Delta_{qQ}^{\prime \dagger} \\
{\bf 0} & M_Q^\dagger \end{array} \right) .
\end{equation}
This matrix has the same form as $ {\cal M}_{\cal Q} $
with $ \Delta_{qQ}^\prime = {\bf 0} $ by replacing the submatrices as
\begin{eqnarray}
M_q & \rightarrow & M_q^\dagger , \
M_Q \rightarrow M_Q^\dagger ,
\nonumber \\
\Delta_{qQ}^\prime = {\bf 0}
& \rightarrow & \Delta_{qQ}^\dagger = {\bf 0} , \
\Delta_{qQ} \rightarrow \Delta_{qQ}^{\prime \dagger} .
\label{eqn:MQMQdagger}
\end{eqnarray}
Then, we have the hermite conjugate of Eq. (\ref{eqn:MQdiagonal}) as
\begin{equation}
{\cal V}_{{\cal Q}_{\rm L}}^\dagger
{\cal M}_{\cal Q}^\dagger {\cal V}_{{\cal Q}_{\rm R}}
= \left( \begin{array}{cc}
{\bar M}_q & {\bf 0} \\ {\bf 0} & {\bar M}_Q \end{array} \right) .
\label{eqn:MQdaggerdiagonal}
\end{equation}
Hence, the relations obtained for the basis (a)
with $ \Delta_{qQ}^\prime = {\bf 0} $ are also applicable
to the present basis (b) with $ \Delta_{qQ} = {\bf 0} $
by taking the substitution (\ref{eqn:MQMQdagger}).
The $ q $-$ Q $ mixing parameters are then replaced as
\begin{equation}
\epsilon^f_i \rightarrow \epsilon^h_i .
\end{equation}
The quark mixing matrices are exchanged as
\begin{equation}
{\cal V}_{{\cal Q}_{\rm L}} \leftrightarrow {\cal V}_{{\cal Q}_{\rm R}} ,
\end{equation}
i.e.,
\begin{eqnarray}
V_{q_{\rm L}} & \leftrightarrow & V_{q_{\rm R}} , \
V_{Q_{\rm L}} \leftrightarrow V_{Q_{\rm R}} ,
\nonumber \\
\epsilon_{q_{\rm L}} & \leftrightarrow & \epsilon_{q_{\rm R}} , \
\epsilon_{q_{\rm L}}^\prime \leftrightarrow \epsilon_{q_{\rm R}}^\prime .
\label{eqn:a-b}
\end{eqnarray}
The ordinary quark masses are given by
\begin{equation}
m_{q_i} = ( \lambda_{q_i} v / {\sqrt 2} )
\left[ 1 + \xi^\prime_{q_i} ( {\mbox{\boldmath $ \epsilon $}}^h )
( \epsilon^h_i )^2 \right]
\label{eqn:mqi-b}
\end{equation}
with certain factors
$ \xi^\prime_{q_i} ( {\mbox{\boldmath $ \epsilon $}}^f ) \sim 1 $,
which are $ - 1/2 $ in the leading order for $ N_Q = 1 $.

\subsection{Seesaw model}

In the seesaw model, the $ q $-$ Q $ mixing matrices
at the first step are obtained
from Eqs. (\ref{eqn:eqL1}) and (\ref{eqn:eqR1}) with $ M_q = {\bf 0} $ as
\begin{eqnarray}
\epsilon_{q_{\rm L}}^{(1)}
&=& \Delta_{qQ}^{\prime \dagger}
( M_Q^\dagger + \delta_{Q_{\rm L}}^\dagger )^{-1} ,
\label{eqn:eqL1-ss}
\\
\epsilon_{q_{\rm R}}^{(1)}
&=& \Delta_{qQ} ( M_Q + \delta_{Q_{\rm R}} )^{-1} .
\label{eqn:eqR1-ss}
\end{eqnarray}
Then, we obtain from Eq. (\ref{eqn:Mq1})
\begin{equation}
M_q^{(1)} = - \Delta_{qQ} ( M_Q + \delta_Q )^{-1}
\Delta_{qQ}^\prime ,
\label{eqn:Mq1-ss}
\end{equation}
where
\begin{eqnarray}
\delta_Q &=&
( C_Q \epsilon_{q_{\rm R}}^{(1) \dagger} \Delta_{qQ} - \delta_{Q_{\rm L}} )
[ {\bf 1} + {\cal E}_Q \epsilon_{q_{\rm R}}^{(1) \dagger} \Delta_{qQ} ]^{-1}
\nonumber \\
& \sim & \Delta_{qQ}^\prime \epsilon_{q_{\rm L}}^{(1)}
+ \epsilon_{q_{\rm R}}^{(1) \dagger} \Delta_{qQ}
\label{eqn:dltQ}
\end{eqnarray}
with
\begin{eqnarray}
C_Q &=& M_Q ( M_Q + \delta_{Q_{\rm R}} )^{-1}
( V_{Q_{\rm R}}^{(1) \dagger -1} - A_{q_{\rm R}} ) ,
\\
{\cal E}_Q &=& ( M_Q + \delta_{Q_{\rm R}} )^{-1}
( V_{Q_{\rm R}}^{(1) \dagger -1} - A_{q_{\rm R}} ) .
\end{eqnarray}

Now suppose that all the singlet quarks have comparable masses,
$ m_{Q_1} , \ m_{Q_2} , \ m_{Q_3} \sim m_Q $.
Then, we obtain from Eqs. (\ref{eqn:eqL1-ss}) and (\ref{eqn:eqR1-ss})
\begin{eqnarray}
( \epsilon_{q_{\rm L}}^{(1)} )_{ia}
& \sim & \epsilon^h_i ,
\label{eqn:eqL1-sss}
\\
( \epsilon_{q_{\rm R}}^{(1)} )_{ia}
& \sim & \epsilon^f_i ,
\label{eqn:eqR1-sss}
\\
( V_{q_{\rm L}}^{(1)} )_{ij}
& \sim & \delta_{ij} + \epsilon^h_i \epsilon^h_j ,
\label{eqn:VqL1-sss}
\\
( V_{q_{\rm R}}^{(1)} )_{ij}
& \sim & \delta_{ij} + \epsilon^f_i \epsilon^f_j .
\label{eqn:VqR1-sss}
\end{eqnarray}
We can also see from Eq. (\ref{eqn:Mq1-ss})
that the effective quark mass matrix has a specific flavor structure,
\begin{equation}
( M_q^{(1)} )_{ij} \sim \epsilon^f_i \epsilon^h_j m_Q .
\label{eqn00:Mq1-sss}
\end{equation}
(The quark mass matrix with this type of flavor structure
is also considered in the context of standard model
with multiple Higgs doublets
by assuming the hierarchy in the Yukawa couplings \cite{HW}.)
The unitary matrices to diagonalize $ M_q^{(1)} $ are given as
\begin{eqnarray}
( V_{q_{\rm L}}^{(2)} )_{ij}
& \sim & \delta_{ij} + \frac{\epsilon^h_i \epsilon^h_j}
{( \epsilon^h_i )^2 + ( \epsilon^h_j )^2} ,
\label{eqn:VqL2-sss}
\\
( V_{q_{\rm R}}^{(2)} )_{ij}
& \sim & \delta_{ij} + \frac{\epsilon^f_i \epsilon^f_j}
{( \epsilon^f_i )^2 + ( \epsilon^f_j )^2} .
\label{eqn:VqR2-sss}
\end{eqnarray}
Then, the net effects on the quark mixings are calculated as
\begin{eqnarray}
\epsilon_{q_\chi} &,& \epsilon_{q_\chi}^\prime
\sim \epsilon^{(1)}_{q_\chi} \ ( \chi = {\rm L} , {\rm R} ) ,
\label{eqn:eqchi-ss} \\
V_{q_\chi} & \sim & V_{q_\chi}^{(2)} \ ( \chi = {\rm L} , {\rm R} ) .
\label{eqn:Vqchi-ss}
\end{eqnarray}
Here, the inequality $ ( V_{q_\chi}^{(1)} )_{ik} ( V_{q_\chi}^{(2)} )_{kj}
\lesssim ( V_{q_\chi}^{(1)} )_{ij} \lesssim ( V_{q_\chi}^{(2)} )_{ij} $
($ k \not= i,j $)
for $ \epsilon^f_i , \epsilon^h_i \lesssim 1 $ is considered.
The ordinary quark masses are obtained as
\begin{equation}
m_{q_i} \sim \epsilon^f_i \epsilon^h_i m_Q .
\label{eqn:mqi-sss}
\end{equation}

We next consider the case with inverted hierarchy
for the singlet quark masses, $ m_{Q_1} \gg m_{Q_2} \gg m_{Q_3} $
\cite{ssinverted}.
The $ q $-$ Q $ mixing terms are assumed to have no significant
flavor dependence, i.e., $ ( \Delta_{qQ} )_{ia} \sim {\bar f} v_S $
and $ ( \Delta_{qQ}^\prime )_{ai} \sim {\bar h} v $.
We may start with
\begin{equation}
M_Q = {\rm diag.} ( m_{Q_1}^0 , m_{Q_2}^0 , m_{Q_3}^0 )
\end{equation}
by using suitable transformations of singlet quarks.
The singlet quark masses are provided dominantly by this $ M_Q $ term
for $ m_{Q_a}^0 \gg {\bar f} v_S , \ {\bar h} v $:
\begin{equation}
m_{Q_a} \simeq m_{Q_a}^0 .
\end{equation}
(This relation is replaced by that of $ m_T \sim {\bar f} v_S $
for the singlet quark $ T \equiv U_3 $
with $ m_T^0 \ll {\bar f} v_S $ \cite{ssinverted}.)
It is also suitable to deform the $ q $-$ Q $ mixing terms
to triangular forms by the ordinary quark transformations,
without modifying $ M_q = {\bf 0} $ and $ M_Q $:
\begin{eqnarray}
\Delta_{qQ} &=& {\bar f} v_S \left( \begin{array}{ccc}
1 & 0 & 0 \\ 1 & 1 & 0 \\ 1 & 1 & 1 \end{array} \right) ,
\label{eqn:DqQ-ssi} \\
\Delta_{qQ}^\prime &=& {\bar h} v \left( \begin{array}{ccc}
1 & 1 & 1 \\ 0 & 1 & 1 \\ 0 & 0 & 1 \end{array} \right) ,
\label{eqn:DqQp-ssi}
\end{eqnarray}
where ``1" denotes the factors of $ O(1) $.

The effective quark mass matrix $ M_q^{(1)} $ given
in Eq. (\ref{eqn:Mq1-ss}) is evaluated as follows.
We first note the identity with the diagonal $ M_Q $,
\begin{eqnarray}
& & ( M_Q + \delta_Q )^{-1}_{ab} = m_{Q_a}^{0 -1} \delta_{ab}
\nonumber \\
& & - m_{Q_a}^{0 -1} \left[ \delta_Q
( {\bf 1} + M_Q^{-1} \delta_Q )^{-1} \right]_{ab} m_{Q_b}^{0 -1} ,
\label{eqn:MQexp}
\end{eqnarray}
where, as seen from Eq. (\ref{eqn:dltQ})
with $ \epsilon_{q_\chi}^{(1)} \lesssim 1 $
and Eqs. (\ref{eqn:DqQ-ssi}) and (\ref{eqn:DqQp-ssi}),
\begin{equation}
\left| \left[ \delta_Q ( {\bf 1} + M_Q^{-1} \delta_Q )^{-1}
\right]_{ab} \right| \lesssim {\bar f} v_S + {\bar h} v .
\end{equation}
Then, by applying Eqs. (\ref{eqn:DqQ-ssi}), (\ref{eqn:DqQp-ssi})
and (\ref{eqn:MQexp}) for Eq. (\ref{eqn:Mq1-ss}),
the effective quark mass matrix $ M_q^{(1)} $ is evaluated
in terms of $ m_{Q_a}^{-1} \simeq m_{Q_a}^{0 -1} $ as
\begin{equation}
M_q^{(1)} \sim ( {\bar h} v ) ( {\bar f} v_S )
\left( \begin{array}{ccc}
m_{Q_1}^{-1} & m_{Q_1}^{-1} & m_{Q_1}^{-1} \\
m_{Q_1}^{-1} & m_{Q_2}^{-1} & m_{Q_2}^{-1} \\
m_{Q_1}^{-1} & m_{Q_2}^{-1} & m_{Q_3}^{-1}
\end{array} \right) .
\label{eqn:Mq1-ssi}
\end{equation}
The second term in the right-hand side of Eq. (\ref{eqn:MQexp}) actually
provides sub-leading contributions as
$ ( {\bar h} v ) ( {\bar f} v_S )^2 m_{Q_i}^{-1} m_{Q_j}^{-1} $.
These corrections, however, do not alter the structure
(\ref{eqn:Mq1-ssi}) of $ M_q^{(1)} $
for $ {\bar f} v_S / m_{Q_i} \lesssim 1 $.
This specific form of $ M_q^{(1)} $ provides the ordinary quark masses,
\begin{equation}
m_{q_i} \sim ( {\bar f} v_S / m_{Q_i} ) {\bar h} v .
\label{eqn:mqi-ssi}
\end{equation}

The first step mixings are obtained
from Eqs. (\ref{eqn:eqL1-ss}) and (\ref{eqn:eqR1-ss})
by using Eqs. (\ref{eqn:DqQ-ssi}) and (\ref{eqn:DqQp-ssi})
for $ \Delta_{qQ} $ and $ \Delta_{qQ}^\prime $
and the relation similar to Eq. (\ref{eqn:MQexp}):
\begin{eqnarray}
( \epsilon_{q_{\rm L}}^{(1)} )_{ia}
& \sim & {\bar h} v / m_{Q_i} ,
\label{eqn:eqL1-ssi}
\\
( \epsilon_{q_{\rm R}}^{(1)} )_{ia}
& \sim & {\bar f} v_S / m_{Q_i} ,
\label{eqn:eqR1-ssi}
\\
( V_{q_{\rm L}}^{(1)} )_{ij}
& \sim & \delta_{ij} + \frac{( {\bar h} v )^2}{m_{Q_i} m_{Q_j}} ,
\label{eqn:VqL1-ssi}
\\
( V_{q_{\rm R}}^{(1)} )_{ij}
& \sim & \delta_{ij} + \frac{( {\bar f} v_S )^2}{m_{Q_i} m_{Q_j}} .
\label{eqn:VqR1-ssi}
\end{eqnarray}
The unitary transformations to diagonalize $ M_q^{(1)} $
of Eq. (\ref{eqn:Mq1-ssi}) are given by
\begin{equation}
( V_{q_{\rm L}}^{(2)} )_{ij} , \ ( V_{q_{\rm R}}^{(2)} )_{ij}
\sim \delta_{ij}
+ \frac{m_{Q_i}^{-1} m_{Q_j}^{-1}}{m_{Q_i}^{-2} + m_{Q_j}^{-2}} .
\label{eqn:VqLR2-ssi}
\end{equation}
It is here noticed that these relations
(\ref{eqn:eqL1-ssi}) -- (\ref{eqn:VqLR2-ssi}) for the quark mixings
are apparently reproduced from
Eqs. (\ref{eqn:eqL1-sss}) -- (\ref{eqn:VqR2-sss})
for the case of $ m_{Q_1} , m_{Q_2} , m_{Q_3} \sim m_Q $
by taking the substitution
\begin{equation}
\epsilon^h_i \rightarrow {\bar h} v / m_{Q_i} , \
\epsilon^f_i \rightarrow {\bar f} v_S / m_{Q_i} .
\end{equation}
Hence, even in this case of $ m_{Q_1} \ll m_{Q_2} \ll m_{Q_3} $,
the net effects on the quark mixings are given
as Eqs. (\ref{eqn:eqchi-ss}) and (\ref{eqn:Vqchi-ss}).
It is interesting that these quark mixings can be expressed
in term of the ordinary quark masses
by using the mass formula (\ref{eqn:mqi-ssi}).

Some remarks should finally be presented in order.
As for the $ t $ and $ T $ quarks
with $ ( M_U )_{33} = m_T^0 \ll {\bar f} v_S $,
the above arguments are still valid
to obtain $ m_t \sim {\bar h} v $ and $ m_T \sim {\bar f} v_S $.
This can be understood by modifying the mass matrix
in Eq. (\ref{eqn:MQexp})
as $ M_U + \delta_U = {\tilde M}_U + {\tilde \delta}_U $
with $ ( {\tilde M}_U )_{33} = ( M_U )_{33} + ( \delta_U )_{33}
\sim {\bar f} v_S $.
In the general quark basis for the seesaw model,
the $ \Delta_{qQ} $ and $ \Delta_{qQ}^\prime $ terms
do not have the triangular forms
(\ref{eqn:DqQ-ssi}) and (\ref{eqn:DqQp-ssi}).
Then, the following substitution should be made
in the above relations for the quark mixings,
\begin{equation}
\epsilon_{q_\chi}
\rightarrow {\tilde V}_{q_\chi}^\dagger \epsilon_{q_\chi} , \
V_{q_\chi} \rightarrow {\tilde V}_{q_\chi}^\dagger V_{q_\chi} .
\end{equation}
Here, $ {\tilde V}_{q_{\rm R}}^\dagger \Delta_{qQ} $
and $ \Delta_{qQ}^\prime {\tilde V}_{q_{\rm L}} $
become the triangular forms (\ref{eqn:DqQ-ssi}) and (\ref{eqn:DqQp-ssi})
by suitably chosing the unitary transformations $ {\tilde V}_{q_\chi} $.

\clearpage


\begin{table}
\caption{
The ordinary quark masses and mixings are listed
for the representative bases,
which are obtained in the presence of singlet quarks.
For the seesaw model,
[com] indicates the case of comparable singlet quark masses
$ m_{Q_1} , m_{Q_2} , m_{Q_3} \sim m_Q $,
and [inv] the case of inverted hierarchy
$ m_{Q_1} \gg m_{Q_2} \gg m_{Q_3} $.
}
\begin{tabular}{llccccc}
\multicolumn{2}{c}{quark mass}
 & ordinary
 & \multicolumn{2}{c}{$ q $-$ Q $ mixings}
 & \multicolumn{2}{c}{ordinary quark mixings} \\
\multicolumn{2}{c}{matrix form}
 & quark masses 
 & [left-handed] & [right-handed]
 & [left-handed] & [right-handed] \\
\multicolumn{2}{c}{$ {\cal M}_{\cal Q} $}
 & $ m_{q_i} $
 & $ ( \epsilon_{q_{\rm L}} )_{ia} $, \
   $ ( \epsilon_{q_{\rm L}}^\prime )_{ia} $
 & $ ( \epsilon_{q_{\rm R}} )_{ia} $, \
   $ ( \epsilon_{q_{\rm R}}^\prime )_{ia} $
 & $ ( V_{q_{\rm L}} )_{ij} (i \not= j) $
 & $ ( V_{q_{\rm R}} )_{ij} (i \not= j) $
\\ \hline
\multicolumn{2}{l}{basis (a) : $ \Delta_{qQ}^\prime = {\bf 0} $}
 & $ \lambda_{q_i} v $
 & $ \displaystyle{\frac{m_{q_i}}{m_Q} \epsilon^f_i} $ & $ \epsilon^f_i $
 & $ \displaystyle{
 {\rule[-3.5ex]{0em}{8ex}}
 \frac{m_{q_i} m_{q_j}}{m_{q_i}^2 + m_{q_j}^2}
                   \epsilon^f_i \epsilon^f_j} $
 & $ \displaystyle{\frac{m_{q_j}^2}{m_{q_i}^2 + m_{q_j}^2}
                   \epsilon^f_i \epsilon^f_j} $
\\ \hline
\multicolumn{2}{l} {basis (b) : $ \Delta_{qQ} = {\bf 0} $}
 & $ \lambda_{q_i} v $
 & $ \epsilon^h_i $
 & $ \displaystyle{\frac{m_{q_i}}{m_Q} \epsilon^h_i} $
 & $ \displaystyle{
 {\rule[-3.5ex]{0em}{8ex}}
 \frac{m_{q_j}^2}{m_{q_i}^2 + m_{q_j}^2}
                   \epsilon^h_i \epsilon^h_j} $
 & $ \displaystyle{\frac{m_{q_i} m_{q_j}}{m_{q_i}^2 + m_{q_j}^2}
                   \epsilon^h_i \epsilon^h_j} $
\\ \hline
seesaw : $ M_q = {\bf 0} $ & [com]
 & $ \epsilon^f_i \epsilon^h_i m_Q $
 & $ \epsilon^h_i $ & $ \epsilon^f_i $
 & $ \displaystyle{
 {\rule[-3.5ex]{0em}{8ex}}
 \frac{\epsilon^h_i \epsilon^h_j}
                        {( \epsilon^h_i )^2 + ( \epsilon^h_j )^2}} $
 & $ \displaystyle{\frac{\epsilon^f_i \epsilon^f_j}
                        {( \epsilon^f_i )^2 + ( \epsilon^f_j )^2}} $
\\
{ } & [inv]
 & $ \displaystyle{\frac{{\bar f} v_S}{m_{Q_i}} {\bar h} v} $
 & $ \displaystyle{\frac{m_{q_i}}{{\bar f} v_S}} $
 & $ \displaystyle{\frac{m_{q_i}}{{\bar h} v}} $
 & $ \displaystyle{
 {\rule[-3.5ex]{0em}{8ex}}
 \frac{m_{q_i} m_{q_j}}{m_{q_i}^2 + m_{q_j}^2}} $
 & $ \displaystyle{\frac{m_{q_i} m_{q_j}}{m_{q_i}^2 + m_{q_j}^2}} $
\end{tabular}
\label{table:mixing-mass}
\end{table}

\mediumtext

\begin{table}
\caption{
The FCNC's in the gauge and scalar interactions
are listed for the representative bases,
which are provided by the $ q $-$ Q $ mixing.
For the seesaw model,
[com] indicates the case of comparable singlet quark masses
$ m_{Q_1} , m_{Q_2} , m_{Q_3} \sim m_Q $,
and [inv] the case of inverted hierarchy
$ m_{Q_1} \gg m_{Q_2} \gg m_{Q_3} $.
}
\begin{tabular}{llcc}
\multicolumn{2}{c}{quark mass}
 & gauge-mediated & scalar-mediated \\
\multicolumn{2}{c}{matrix form}
 & FCNC's ($ i \not= j $)
 & FCNC's ($ i \not= j $) \\
\multicolumn{2}{c}{$ {\cal M}_{\cal Q} $}
 & $ \Delta {\cal Z}_{\cal Q} [q]_{ij} $
 & $ \Lambda_{\cal Q}^\alpha [q]_{ij} $
\\ \hline
\multicolumn{2}{l}{basis (a) : $ \Delta_{qQ}^\prime = {\bf 0} $}
 & $ \displaystyle{\frac{m_{q_i}}{m_Q} \frac{m_{q_j}}{m_Q}
   \epsilon^f_i \epsilon^f_j} $
 & $ \displaystyle{
 {\rule[-3ex]{0em}{6.5ex}}
 \frac{m_{q_j}}{v_S} \epsilon^f_i \epsilon^f_j
   + \frac{m_{q_i}}{v} \Delta {\cal Z}_{\cal Q} [q]_{ij}} $
\\ \hline
\multicolumn{2}{l}{basis (b) : $ \Delta_{qQ} = {\bf 0} $}
 & $ \epsilon^h_i \epsilon^h_j $
 & $ \displaystyle{
 {\rule[-3ex]{0em}{6.5ex}}
 \frac{m_Q}{v_S} {\bar f}_i \epsilon^h_j
   + \frac{m_{q_i}}{v} \Delta {\cal Z}_{\cal Q} [q]_{ij}} $
\\ \hline
seesaw : $ M_q = {\bf 0} $ & [com]
 & $ \epsilon^h_i \epsilon^h_j $
 & $ \displaystyle{
 {\rule[-3ex]{0em}{6.5ex}}
 \frac{m_{q_j}}{v_S} \epsilon^f_i \epsilon^f_j
   + \frac{m_{q_i}}{v} \Delta {\cal Z}_{\cal Q} [q]_{ij}} $
\\
{ } & [inv]
 & $ \displaystyle{\frac{m_{q_i}}{{\bar f} v_S}
   \frac{m_{q_j}}{{\bar f} v_S}} $
 & $ \displaystyle{
 {\rule[-3ex]{0em}{6.5ex}}
 \frac{m_{q_i}}{{\bar h} v} \frac{m_{q_j}}{{\bar h} v}
   \frac{m_{q_j}}{v_S}
   + \frac{m_{q_i}}{v} \Delta {\cal Z}_{\cal Q} [q]_{ij}} $
\end{tabular}
\label{table:FCNC's}
\end{table}

\clearpage

\vspace*{1cm}

\clearpage

\vspace*{1cm}

\clearpage

\vspace*{1cm}

\clearpage

\vspace*{1cm}

\clearpage

\vspace*{1cm}

\clearpage

\vspace*{1cm}

\clearpage

\vspace*{1cm}

\end{document}